\documentclass[conference]{IEEEtran}
\IEEEoverridecommandlockouts
\usepackage{cite}
\usepackage{amsmath,amssymb,amsfonts}
\usepackage{booktabs}
\usepackage{algorithmic}
\usepackage{graphicx}
\usepackage{textcomp}
\usepackage{algorithm}
\def\BibTeX{{\rm B\kern-.05em{\sc i\kern-.025em b}\kern-.08em
    T\kern-.1667em\lower.7ex\hbox{E}\kern-.125emX}}
\usepackage{xspace}
\usepackage{multirow}
\usepackage{array}
\usepackage[dvipsnames,table,xcdraw]{xcolor}
\usepackage[hidelinks]{hyperref}
\usepackage{url}
\usepackage{breakurl}
 
\usepackage[caption=false]{subfig}

\usepackage{pifont}
\usepackage{pgfplots}
\usepgfplotslibrary{statistics}
\usepackage{pgfplotstable}
\pgfplotsset{compat=newest}
\usepackage{filecontents}
\usepackage{lipsum}
\usepackage{soul}
\usepackage{comment}
\usepackage{footnote}
\usepackage{siunitx}
\usepackage{etoolbox}
\usepackage{soulpos}
\usepackage{wrapfig}
\usepackage{siunitx}
\usepackage{atbegshi}
\usepackage{etoolbox}

\usepackage{siunitx}
\usepackage{enumitem}

\hypersetup{colorlinks=true,
citecolor=Maroon,
linkcolor=Green,
urlcolor=Maroon,
breaklinks=true,
}

\makesavenoteenv{tabular}
\makesavenoteenv{table}

\def\BibTeX{{\rm B\kern-.05em{\sc i\kern-.025em b}\kern-.08em
    T\kern-.1667em\lower.7ex\hbox{E}\kern-.125emX}}

\ulposdef{\gray}[xoffset=1pt]{\mbox{\color{gray!30}\rule[-.8ex]{\ulwidth}{3ex}}}

\newcommand{\eg}{\textit{e.g.}}

\newcommand{\etal}{\textit{et al.}}

\newcommand{\orin}{\textsc{Orin-Nano}\xspace}
\newcommand{\xavier}{\textsc{Xavier-NX}\xspace}

\newcommand{\sysname}{\textsc{Coral}\xspace}
\newcommand{\yolo}{\textsc{YOLO}\xspace}
\newcommand{\frcnn}{\textsc{FRCNN}\xspace}
\newcommand{\retina}{\textsc{RetinaNet}\xspace}

\def\footnoterule{\relax
  \kern0pt
  \hbox to \columnwidth{\hfill\vrule width 1.0\columnwidth height 0.4pt\hfill}
  \kern4.6pt}

\newcommand{\para}[1]{\vspace{1mm}\noindent{\bf #1.}}

\begin{document}

\title{\huge Covariance-Guided Resource Adaptive Learning for Efficient Edge Inference}

\author{\IEEEauthorblockN{
    \parbox{\linewidth}{\centering
    Ahmad N. L. Nabhaan\IEEEauthorrefmark{1}\textsuperscript{\textsection},
    Zaki Sukma\IEEEauthorrefmark{2}\textsuperscript{\textsection}, 
    Rakandhiya D. Rachmanto\IEEEauthorrefmark{2}, \\
    Muhammad Husni Santriaji\IEEEauthorrefmark{3},
    Byungjin Cho\IEEEauthorrefmark{5},
    Arief Setyanto\IEEEauthorrefmark{1},
    In Kee Kim\IEEEauthorrefmark{2}}}
    \IEEEauthorblockA{\IEEEauthorrefmark{1}Universitas Amikom Yogyakarta, Department of Informatics, \{ahmad.nabhaan, arief\_s\}@amikom.ac.id}
    \IEEEauthorblockA{\IEEEauthorrefmark{2}University of Georgia, School of Computing, \{zaki.sukma, rakandhiya.rachmanto, inkee.kim\}@uga.edu}

    \IEEEauthorblockA{\IEEEauthorrefmark{3}Universitas Gadjah Mada, Department of Computer Science and Electronics, muhammad.husnisantriaji@ugm.ac.id}

    \IEEEauthorblockA{\IEEEauthorrefmark{5}Korea University, Department of Computer Science and Software Engineering, byungjincho@korea.ac.kr}
}

\pagenumbering{arabic}
\maketitle
\thispagestyle{plain}
\pagestyle{plain}

\begingroup
\renewcommand\thefootnote{\textsection}
\footnotetext{Nabhaan and Sukma contributed equally and are co-first authors.}
\endgroup

\begingroup\renewcommand\thefootnote{\textsection}
\endgroup

\begin{abstract}
% Real-world deployments of edge AI demand consistent inference throughput under strict power constraints, a balance difficult to achieve due to the complex, non-linear interactions of hardware parameters. Existing solutions often rely on inefficient static presets or require expensive, time-consuming offline profiling. To address this, we present CORAL (Correlation-Guided Resource Adaptive Learning), an online optimization framework that discovers near-optimal configurations without offline profiling. CORAL leverages distance covariance to statistically capture the non-linear dependencies between hardware settings—such as DVFS and concurrency levels—and performance metrics. Unlike prior work, we explicitly formulate the challenge as a dual-target optimization problem to satisfy power limits and throughput targets simultaneously. We evaluate CORAL on NVIDIA Jetson Xavier NX and Orin Nano devices across three object detection models ranging from lightweight to heavyweight. In single-target scenarios, CORAL achieves 96–100\% of the optimal performance found by exhaustive search. In a strict dual-target scenarios where baselines fail or exceed power budgets, CORAL successfully converges to valid configurations within several iterations.
For deep learning inference on edge devices, hardware configurations achieving the same throughput can differ by 2$\times$ in power consumption, yet operators often struggle to find the efficient ones without exhaustive profiling. 
Existing approaches often rely on inefficient static presets or require expensive offline profiling that must be repeated for each new model or device. To address this problem, we present \sysname, an online optimization method that discovers near-optimal configurations without offline profiling. 
\sysname leverages distance covariance to statistically capture the non-linear dependencies between hardware settings, \eg, DVFS and concurrency levels, and performance metrics. 
Unlike prior work, we explicitly formulate the challenge as a throughput-power co-optimization problem to satisfy power budgets and throughput targets simultaneously. 
We evaluate \sysname on two NVIDIA Jetson devices across three object detection models ranging from lightweight to heavyweight. 
In single-target scenarios, \sysname achieves 96\% -- 100\% of the optimal performance found by exhaustive search. 
In strict dual-constraint scenarios where baselines fail or exceed power budgets, \sysname consistently finds proper configurations online with minimal exploration.

\end{abstract}


\section{Introduction}\label{sec:intro}
Real-world deployments of edge applications, \eg, from autonomous drones to smart surveillance, demand consistent inference throughput under strict power constraints \cite{edgeai.pieee23}. 
For example, a drone running object detection must sustain a sufficient frame rate for responsive navigation while staying within a power budget that preserves battery life (and sufficient flying time as well). 
Both constraints -- {\em throughput} and {\em power requirements} -- must be satisfied simultaneously. However, achieving this balance is a non-trivial research problem \cite{convergo.edge25}.

The core difficulty lies in the sheer number of possible configurations on modern edge hardware. 
Edge devices like NVIDIA Jetson have dozens of tunable parameters, including CPU and GPU frequencies, memory clock rates, and active core counts \cite{dvfo2024, polythrottle2024}. 
However, interactions among these parameters are complex and exhibit non-linear behavior. 
Changing one parameter affects the impact of others, making it difficult to predict performance from individual settings alone. 
Manufacturers simplify this complexity by offering a handful of preset power modes, but these presets represent only a tiny sample of the possible configurations \cite{edgeptq.edge24}. 
In battery-constrained deployments, {\tt max-power} presets can consume excessive power. 
For real-time applications, {\tt power-saving} presets may deliver too few frames. 
Therefore, operators and edge practitioners are often asked to choose between excessive power draw and insufficient throughput.

\begin{figure}[t]
    \centering
    \includegraphics[width=0.85\linewidth]{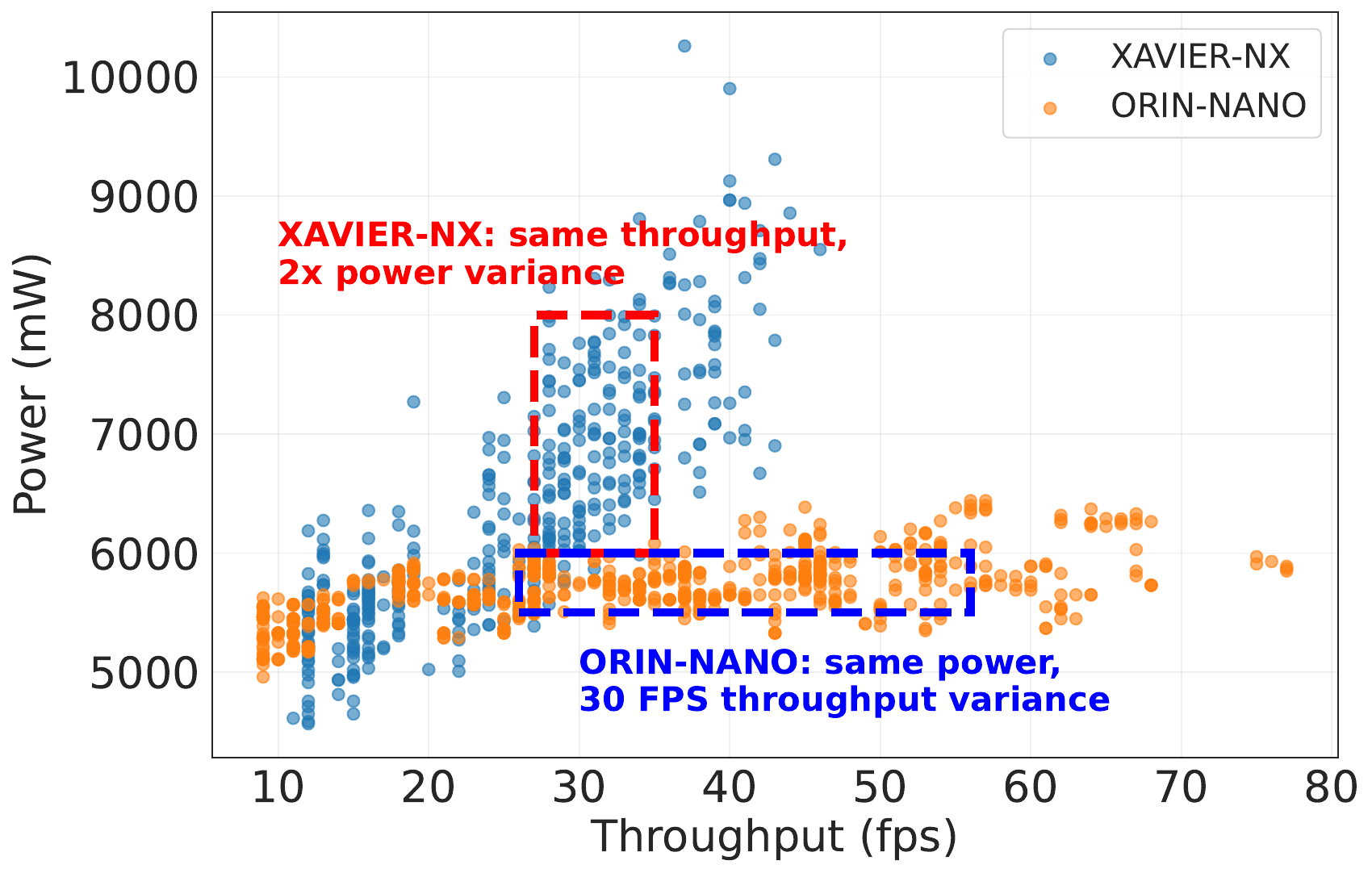}
    \caption{{\bf Power-throughput trade-off of YOLOv5-N on Xavier NX and Orin Nano.} {\em The \textcolor{black}{red box (\xavier)} highlights configurations with similar throughput but varying power consumption; the \textcolor{black}{blue box (\orin)} shows configurations achieving different throughputs at similar power levels.}}
    \label{fig:scatter}
\end{figure}

Our profiling results confirm this difficulty. 
Figure \ref{fig:scatter} shows the power-throughput trade-off of YOLOv5-N on two Jetson devices. 
On Xavier NX, configurations achieving similar throughput ($\sim$30 fps) exhibit up to 2$\times$ difference in power consumption (6\si{\watt}--8\si{\watt}).
Conversely, Orin Nano demonstrates that comparable power levels ($\sim$6\si{\watt}) can yield throughputs ranging from 40 to 75 fps.
These results show that configuration choice significantly impacts efficiency. 
Poor choices can double power consumption, while others achieve high throughput at minimal power. Finding these efficient operating points without exhaustive search is the key challenge.

Several approaches have been proposed, but each has its own limitations. 
Profiling-based methods, \eg, ALERT \cite{alert.atc20}, can achieve near-optimal performance but require exhaustive offline profiling. 
This process must be repeated for each new model or device, consuming hours of profiling time and energy. Learning-based methods often use deep reinforcement learning \cite{dvfo2024} that adapt online but require extensive training time and many trial-and-error iterations to converge. 
Moreover, both approaches typically optimize for a single metric, either throughput or power, leaving dual-constraint scenarios largely unaddressed. The irregular search space, where many configurations satisfy one constraint but violate the other, defeats conventional optimization strategies \cite{runtimeohedge.ispa21}.

We present \sysname (\underline{Co}variance-Guided \underline{R}esource \underline{A}daptive \underline{L}earning), an online optimization method that discovers near-optimal configurations on edge devices by exploiting the following key insight.
{The statistical relationship between hardware setting changes and their performance impact can be captured through distance covariance, which does not require exhaustive profiling.}

Distance covariance \cite{szekely2009brownian} measures dependency between variables regardless of their relationship's functional form, making it ideal for the non-linear configuration space of edge devices. 
\sysname computes correlation weights between each hardware parameter (CPU frequency, GPU frequency, memory clock, core count, concurrency level) and both target metrics (throughput and power). 
These weights guide a lightweight search that prioritizes high-impact parameters, converging to valid configurations within 10 iterations, orders of magnitude faster than profiling-based alternatives. 
Unlike prior work \cite{alert.atc20, kalmia.infocom22} that optimizes for throughput alone or power alone, \sysname explicitly formulates the problem as throughput-power co-optimization. 
By maintaining separate correlation weights for throughput and power, \sysname identifies configurations that satisfy both constraints simultaneously.

\begin{table}[t]
\caption{{\bf Hardware Specifications of NVIDIA Jetson Devices.} \em }
\centering
\begin{tabular}{l|l|l}
\toprule 
 & \textbf{\xavier} \cite{xaviernx.web24} & \textbf{\orin} \cite{orinnano.web24} \\ \midrule
\textbf{CPU} &
  \begin{tabular}[c]{@{}l@{}}6-core Carmel \\ @1.9GHz\end{tabular} &
  \begin{tabular}[c]{@{}l@{}}6-core Cortex-A78AE \\ @1.5GHz\end{tabular} \\ \midrule 
\textbf{GPU} &
  \begin{tabular}[c]{@{}l@{}}384-core Volta,
  \\ 48 Tensor cores
  \\ @1100 MHz\end{tabular} & \begin{tabular}[c]{@{}l@{}}1024-core Ampere, \\ 32 Tensor cores \\ @625 MHz\end{tabular} \\ \midrule
\textbf{Memory}     & \begin{tabular}[c]{@{}l@{}} Shared 8GB \\ LPDDR4X \end{tabular}       & \begin{tabular}[c]{@{}l@{}} Shared 8GB \\ LPDDR4X \end{tabular}       \\ \midrule
\textbf{OS}         & Ubuntu 20.04              & Ubuntu 22.04              \\ \midrule
\textbf{JetPack}    & ver. 5.1                  & ver 6.1                   \\ \bottomrule
\end{tabular}
\label{tab:devices}
\end{table}

We evaluate \sysname on NVIDIA Jetson Xavier NX and Orin Nano across three object detection models: YOLOv5-N, FRCNN-MobileNetV3, and RetinaNet-ResNet50. 
We compare against four state-of-the-art baselines:  
ORACLE (exhaustive offline profiling of all configurations), ALERT (offline profiling with Kalman filtering) \cite{alert.atc20}, ALERT-Online (online adaptation of ALERT), and manufacturer presets ({\tt max-power} and {\tt default} modes).
Our evaluation results show that when optimizing for throughput alone, \sysname achieves 96\% -- 100\% of ORACLE performance. 
The more challenging case is optimizing for both power and throughput simultaneously. 
Under a 6.5\si{\watt} power limit and 30 fps throughput target on Xavier NX with YOLOv5-N, \sysname achieves 34 fps at 5.9\si{\watt}, satisfying both constraints. 
In contrast, ALERT exceeds the power budget at 8.5\si{\watt}, ALERT-Online fails to find valid configurations, and the default preset delivers only 15 fps. 
These results hold across all model sizes and both devices.

This paper makes the following contributions:

\begin{itemize}
	\item {\bf Online Optimization Without Offline Profiling:} 
	We introduce the use of distance covariance for hardware configuration search, enabling real-time adaptation without offline profiling or expensive model training. To our knowledge, \sysname is the first method to apply this statistical technique to edge inference optimization.

	\item {\bf Throughput-Power Co-Optimization Formulation:} 
	We formalize edge inference as a throughput-power co-optimization problem with explicit power limits and throughput targets. This addresses real deployment constraints but has been largely ignored by prior work.

	\item {\bf Comprehensive Evaluation:} 
	We evaluate \sysname on two real-world NVIDIA Jetson devices across three object detection models spanning 20$\times$ size difference, demonstrating consistent improvements over baselines in both single-metric and dual-constraint scenarios. 
 
\end{itemize}

\section{Background and Related Work}\label{sec:bg-related}

\subsection{Background}
\label{subsec:background}

\subsubsection{Hardware Configuration on Edge Devices}
\label{subsubsec:devices}

Modern edge AI platforms provide fine-grained control over hardware settings through Dynamic Voltage and Frequency Scaling (DVFS) \cite{dvfo2024}. 
NVIDIA Jetson devices exemplify this capability. 
Jetson Xavier NX (\xavier) \cite{xaviernx.web24}, built on the Volta architecture, features six Carmel CPU cores, a 384-core GPU with 48 Tensor cores, and 8GB LPDDR4X memory. 
The newer Jetson Orin Nano (\orin) \cite{orinnano.web24} uses the Ampere architecture with six Cortex-A78AE cores, a 1024-core GPU with 32 Tensor cores, and 8GB LPDDR5 memory. 
Table \ref{tab:devices} summarizes their specifications.

On these platforms, operators can adjust multiple parameters that directly affect both performance and power consumption:

\begin{itemize}
	\item {\bf CPU frequency and core count:} Higher frequencies and more active cores increase computational throughput but also raise power draw.

	\item {\bf GPU frequency:} Important configuration for DL inference, as most neural network operations execute on the GPU.

	\item {\bf Memory frequency:} This configuration affects data transfer rates between memory and processing units.

	\item {\bf Concurrency level:} The number of inference instances running simultaneously \cite{aimultitenancy.cloud21, reachingforsky.tiot23}. Higher concurrency can improve hardware utilization by overlapping CPU preprocessing with GPU computation, but also increases resource contention and power consumption.

\end{itemize}

\begin{table}[t]
    \caption{\textbf{Types and Ranges of Tunable Hardware Parameters.}}
    \centering
    \resizebox{0.95\columnwidth}{!}{
        \begin{tabular}
            {l|c|c}
            \toprule 
            & \textbf{\xavier} \cite{xaviernx.web24} & \textbf{\orin} \cite{orinnano.web24} \\
            \midrule
            CPU Cores & 2 - 6 & 2 - 6 \\
            CPU Frequency & 1190 - 1908 MHz & 806 - 1510 MHz \\
            GPU Frequency & 510 - 1100 MHz & 306 - 624 MHz \\
            Memory Frequency & 1500 - 1866 MHz & 2133 MHz, 3199 MHz \\
            Concurrency Level & 1 - 3 & 1 - 5 \\
            \bottomrule
        \end{tabular}
    }
    \label{tab:dvfs}
    \vspace{-1em}
\end{table}

Table \ref{tab:dvfs} summarizes the tunable parameter ranges, including concurrency levels of 1--3 on Xavier NX and 1--5 on Orin Nano. 
The combination of these parameters creates thousands of possible configurations, each with different power-throughput characteristics.

Manufacturers provide preset power modes (\eg, 10W, 15W) that fix DVFS parameters to predefined values. 
These presets are convenient; however, they cover only a small fraction of the configuration space and may miss efficient operating points. 
Furthermore, they do not consider application-oriented configurations (\eg, concurrency level), leaving additional optimization potential {\em unexplored}.

\subsubsection{Distance Covariance}
\label{secsec:discov}

Distance covariance \cite{szekely2009brownian} measures statistical dependency between two variables without assuming a specific functional form. 
Unlike Pearson correlation, which captures only linear relationships, {\em distance covariance detects arbitrary non-linear dependencies}.
Given $n$ paired observations of metric $m$ (e.g., throughput or power) and hardware setting $s$, we first compute the pairwise distance matrices:
\begin{equation}
a_{ij} = \|m_i - m_j\|, \quad b_{ij} = \|s_i - s_j\|
\end{equation}

These matrices are then double-centered to remove marginal effects:
\begin{equation}
A_{ij} = a_{ij} - \bar{a}_{i\cdot} - \bar{a}_{\cdot j} + \bar{a}_{\cdot\cdot}
\end{equation}
where $\bar{a}_{i\cdot}$, $\bar{a}_{\cdot j}$, and $\bar{a}_{\cdot\cdot}$ denote the row mean, column mean, and grand mean, respectively. The distance covariance is then:
\begin{equation}
\text{dCov}^2(m, s) = \frac{1}{n^2} \sum_{i,j} A_{ij} B_{ij}
\end{equation}

Finally, distance correlation normalizes by the individual variances:
\begin{equation}
\label{eq:corr}
\text{dCor}(m, s) = \frac{\text{dCov}(m, s)}{\sqrt{\text{dCov}(m, m) \cdot \text{dCov}(s, s)}}
\end{equation}

The resulting $\text{dCor}$ ranges from 0 to 1, where 0 indicates statistical independence and higher values indicate stronger dependency. 
Throughout this paper, we use {\em distance covariance} to refer to this methodology, while the algorithm uses the normalized form ($\text{dCor}$) for comparability across parameters. 
This property makes distance covariance well-suited for capturing the complex relationships between hardware settings and performance metrics. The relationship between hardware settings and performance is highly non-linear.
Increasing GPU frequency may boost throughput only when CPU frequency is sufficiently high.

\subsection{Related Work}

\subsubsection{Hardware-Aware Optimization}

A substantial body of work addresses hardware efficiency for DL inference. 
DVFS and power management strategies have been extensively studied to navigate the fundamental trade-off between power and performance.

\noindent
{\bf Profiling-based methods.} 
ALERT \cite{alert.atc20} demonstrated the effectiveness of coordinating application and system-level adaptation, achieving significant energy savings through probabilistic modeling with Kalman filters. 
However, ALERT requires extensive offline profiling that must be repeated for each model-device combination.

\noindent
{\bf Learning-based methods.} 
Zhang \etal \cite{dvfo2024} proposed DVFO, using deep reinforcement learning (DRL) to co-optimize DVFS parameters for edge-cloud collaborative inference. 
While effective, DRL-based methods require extensive training iterations and may struggle to converge quickly in new environments.
Yan \etal \cite{polythrottle2024} introduced PolyThrottle, applying constrained Bayesian optimization for energy-efficient inference, but such methods may still need offline exploration phases.

\noindent
{\bf Online (or offline profiling-free) methods.} SHEEO~\cite{sheeo2022} uses reinforcement learning to continuously optimize energy efficiency without prior profiling, monitoring runtime variations, and adapting power management strategies accordingly.

\sysname differs from these approaches by using distance covariance for lightweight online search, requiring no offline profiling, no training phase, and explicitly targeting dual-constraint scenarios where both power and throughput must be satisfied.

\subsubsection{Concurrent Execution Frameworks}

Recent frameworks address execution orchestration for edge inference. Miriam \cite{miriam2023} proposes a contention-aware task coordination framework for multi-DNN inference on edge GPUs, achieving significant throughput improvements while maintaining low latency for critical tasks. 
HaX-CoNN \cite{haxconn2023} enables heterogeneity-aware execution of concurrent DNNs by intelligently scheduling workloads across diverse accelerators within a System-on-Chip. 
Hao et al. \cite{reachingforsky.tiot23} explored maximizing inference throughput by leveraging AI multi-tenancy and dynamic model placement on heterogeneous edge devices.
\sysname complements these frameworks by functioning as a device-level optimizer. While frameworks like HaX-CoNN manage distributed scheduling, \sysname dynamically tunes the underlying hardware parameters (concurrency level and DVFS settings) in real-time to meet specific throughput and power targets.

\section{The \sysname Method}\label{sec:approach}
\subsection{Problem Formulation}

We aim to find a hardware configuration that satisfies both throughput and power constraints. 
We define the configuration space as:
\begin{equation}
\mathbf{s} = (s_{cpu}, c_{cpu}, s_{gpu}, s_{mem}, c)
\end{equation}
where $s_{cpu}$ is CPU frequency, $c_{cpu}$ is the number of active CPU cores, $s_{gpu}$ is GPU frequency, $s_{mem}$ is memory frequency, and $c$ is concurrency level. Each configuration $\mathbf{s}$ yields a throughput $\tau(\mathbf{s})$ and power consumption $p(\mathbf{s})$.

\para{Throughput-Power Co-Optimization}
Given a throughput target $\tau_{target}$ and power limit $p_{budget}$, we seek a configuration $\mathbf{s}^*$ that satisfies both constraints:
\begin{equation}
\tau(\mathbf{s}^*) \geq \tau_{target} \quad \text{and} \quad p(\mathbf{s}^*) \leq p_{budget}
\end{equation}
A configuration that violates either constraint is marked as infeasible and excluded from future exploration. Among all feasible configurations, we prefer those with higher efficiency $\eta(\mathbf{s}) = \tau(\mathbf{s}) / p(\mathbf{s})$.

\para{Search Strategy}
\sysname adapts its search direction based on the current configuration's performance. 
If the throughput target is already satisfied, \sysname attempts to reduce power consumption to improve efficiency. 
Otherwise, \sysname prioritizes increasing throughput to first reach the feasible region. 
This adaptive strategy enables \sysname to navigate toward the feasible region while optimizing for efficiency.

\para{Challenges} This problem has three challenges. 
First, the configuration space can be large (over 2K combinations per device in our experiments). 
Second, the functions $\tau(\mathbf{s})$ and $p(\mathbf{s})$ are non-linear and exhibit complex interactions among parameters as discussed in \S\ref{sec:intro}. 
Third, the feasible region satisfying both constraints is often 
narrow, making exhaustive search impractical.
These challenges motivate \sysname's distance covariance-based approach, which efficiently identifies high-impact parameters without exhaustive profiling.

% ===================================================
% ===================================================

\subsection{Overview}

\sysname operates in three steps at each iteration.

\begin{description}
% \textbf{Step 1: Reward Evaluation (\S\ref{subsec:reward}).}  

\item \textbf{Step 1: Reward Evaluation (\S\ref{subsec:reward}).}
\sysname executes inference with the current configuration and measures throughput and power. 
It then evaluates the results using Algorithm \ref{alg:calculate_reward} to find 1) configurations satisfying both constraints receive a positive reward (efficiency) and 2) those violating constraints receive a penalty and are added to a prohibited list. 
The best and second-best configurations are updated based on the reward.

% \textbf{Step 2: Correlation Analysis (\S\ref{subsec:corr_analysis}).} 
\item \textbf{Step 2: Correlation Analysis (\S\ref{subsec:corr_analysis}).} 
\sysname computes distance correlation between each hardware parameter and both performance metrics (throughput and power) using recent observations. 
This identifies which parameters have the strongest influence on performance.

% \textbf{Step 3: Configuration Search (\S\ref{subsec:config_search}).} 
\item \textbf{Step 3: Configuration Search (\S\ref{subsec:config_search}).} 
\sysname uses the correlation weights to determine the next configuration to explore. 
Parameters with higher correlation receive larger adjustments, while less influential parameters change minimally (Algorithm \ref{alg:d2cov}).

\end{description}

These steps repeat for a fixed iteration budget (10 in our implementation).

% ===================================================
% ===================================================

\subsection{Reward and Feasibility Check}
\label{subsec:reward}

Algorithm \ref{alg:calculate_reward} evaluates each configuration by checking feasibility and computing a reward score.

\para{Feasibility Check} 
A configuration is feasible if it satisfies both constraints: throughput meets the target ($\tau \geq \tau_{target}$) and power stays within the budget ($p \leq p_{budget}$). 
Configurations violating either constraint are marked as infeasible and added to a prohibited list ($PS$). 
Before exploring a new configuration, \sysname checks $PS$ and skips any previously identified infeasible configurations to avoid redundant evaluations.

\para{Reward Calculation} 
For feasible configurations, the reward is the efficiency ratio:
\begin{equation}
r = \frac{\tau}{p}
\end{equation}

Higher efficiency means more throughput per watt, which is desirable for energy-constrained edge deployments. For infeasible configurations, the algorithm returns a negative penalty:
\begin{equation}
r = -(\frac{p}{\tau})
\end{equation}

We use this penalty to ensure that infeasible configurations always rank below feasible ones. The inverted ratio (power over throughput) further penalizes configurations that waste power without delivering throughput.

\begin{algorithm}[t]
\caption{Reward Calculation}
\label{alg:calculate_reward}
\begin{algorithmic}[1]
\REQUIRE $\mathbf{m}$, $\mathbf{x}$, $PS$ \quad (measured metrics, current setting, prohibited setting list)
\ENSURE $r$ \quad (reward score value)
\STATE $p \gets m_0[\text{power consumption}]$
\STATE $t \gets m_0[\text{throughput}]$
\IF{$t < \tau_{\mathit{target}}$ or $p > p_{\mathit{budget}}$}
    \STATE $PS.\textsc{append}(x)$
    \RETURN $- \left(\frac{p}{t}\right)$ % \times (1 \times 10^{-6})
\ELSE
    \STATE $r \gets \left(\frac{t}{p} \right)$
\ENDIF
\RETURN $r$
\end{algorithmic}
\end{algorithm}

% ===================================================
% ===================================================

\subsection{Correlation Analysis}
\label{subsec:corr_analysis}

\sysname uses distance correlation (Eq-\ref{eq:corr}) to identify which hardware parameters most strongly influence performance. For each parameter dimension $i$, we compute two correlation weights:
\begin{equation}
\alpha_i = dCor(\tau, s_i) \quad \text{and} \quad \beta_i = dCor(p, s_i)
\end{equation}
where $\alpha_i$ measures the correlation with throughput and $\beta_i$ measures the correlation with power. The distance correlation function $dCor(\cdot, \cdot)$ is defined in \S\ref{secsec:discov}, which is to capture non-linear dependencies between parameters, making it suitable for the complex interactions in hardware configuration spaces.

\para{Sliding Window} 
\sysname maintains a sliding window of the $W$ most recent observations. Each observation is a tuple $(s, \tau, p)$ containing the configuration and its measured metrics. Using only recent samples keeps the computation lightweight and allows \sysname to adapt as the search progresses.

To illustrate, suppose the sliding window contains observations with throughput $\tau = [15.2, 16.1, 15.8, 14.9, 15.5]$ fps, power consumption $p = [9800, 10100, 10050, 9500, 9750]$ m\si{\watt}, and CPU frequency settings $s_{cpu} = [1200, 1400, 1400,$ $1000, 1200]$ MHz. 
Computing distance correlation yields $\alpha_{cpu}$ $= 0.94$ and $\beta_{cpu} = 0.99$. 
These high values (close to 1) indicate that CPU frequency strongly influences both throughput and power, with power correlation being slightly higher.

\para{Interpretation} 
A high $\alpha_i$ indicates that parameter $i$ strongly affects throughput; a high $\beta_i$ indicates strong influence on power. In the decision phase, \sysname takes the dominant correlation $\gamma_i = \max(\alpha_i, \beta_i)$ and uses it to scale adjustments,
such that parameters with higher correlation receive larger adjustments, while less influential parameters change minimally.

% ===================================================
% ===================================================

\subsection{Configuration Search}
\label{subsec:config_search}

\begin{algorithm}[t]
\caption{Configuration Search}
\label{alg:d2cov}
\begin{algorithmic}[1]
\REQUIRE 
  $\mathbf{x}$, $\mathbf{y}$, $\boldsymbol{\alpha}$, $\boldsymbol{\beta}$, $\boldsymbol{\tau}$, 
  $\mathbf{r}$, $\tau_{\mathit{target}}$, $\mathit{aside}$, $p$ \quad 
  (best setting, second best setting, throughput correlation, power correlation, throughput, ranges, throughput target, aside flag, power consumption)
\ENSURE $\mathbf{z}$ \quad (next setting)
\STATE $\mathbf{z} \gets [\ ]$

\FOR{each $(x_i, y_i, \alpha_i, \beta_i, \tau_i, r_i)$ in $(\mathbf{x}, \mathbf{y}, \boldsymbol{\alpha}, \boldsymbol{\beta}, \boldsymbol{\tau}, \mathbf{r})$}
    \STATE $\gamma_i \gets \max(\alpha_i, \beta_i)$
    \STATE $\Delta_i \gets \frac{1}{2}|x_i - y_i| \cdot \gamma_i$
    \STATE $(l, h) \gets (y_i, x_i)$ if $\mathit{aside}$ else $(x_i, y_i)$
    \IF{$\tau_{\mathit{last}} > \tau_{\mathit{target}}$ \AND $p_{\mathit{last}} >= p_{\mathit{min}}$}
        \STATE $v_i \gets l - \Delta_i$
    \ELSE
        \STATE $v_i \gets h + \Delta_i$
    \ENDIF
    \STATE $z_i \gets \textsc{minmax}(\textsc{round}(v_i), r_i)$
    \STATE $\mathbf{z}.\textsc{append}(z_i)$
\ENDFOR
\IF{$p_{\mathit{best}} > p_{\mathit{min}}$ \AND $\tau_{\mathit{best}} > \tau_{\mathit{target}}$}
    \STATE $z_{CPU\ Cores} \gets \textsc{min}(\boldsymbol{r}_{CPU\ Cores})$
    \STATE $z_{Concurrency\ Level} \gets \textsc{max}(\boldsymbol{r}_{Concurrency\ Level})$
\ENDIF
\RETURN $\mathbf{z}$

\end{algorithmic}
\end{algorithm}

Algorithm~\ref{alg:d2cov} generates the next configuration based on correlation weights and the current best configurations.

\para{Step Size} 
For each parameter dimension $i$, \sysname computes a combined weight $\gamma_i = \max(\alpha_i, \beta_i)$ and calculates a weighted step size:
\begin{equation}
\Delta_i = \frac{1}{2} |x_i - y_i| \cdot \gamma_i
\end{equation}
where $x_i$ and $y_i$ are the values of parameter $i$ in the best and second-best configurations, respectively. This step size is proportional to both the distance between good configurations and the correlation strength. Parameters with higher correlation receive larger adjustments, while less influential parameters change minimally.

\para{Search Direction} 
\sysname sets low and high bounds ($l$, $h$) from the best and second-best configurations, and adapts its search direction based on current performance. If throughput meets the target and power is above the floor, \sysname moves toward lower values ($v_i = l - \Delta_i$) to reduce power consumption. Otherwise, \sysname moves toward higher values ($v_i = h + \Delta_i$) to improve throughput. After computing each parameter value, \sysname clamps it to the valid range to ensure the configuration is executable.

\para{Power Optimization Heuristic} 
When the throughput target is satisfied but power remains above the floor, \sysname sets CPU frequency to minimum and concurrency to maximum. This leverages the insight that CPU is a dominant power consumer, and higher concurrency can compensate for reduced CPU performance.

\section{Evaluations}\label{sec:eval}

\subsection{Experimental Setup}
We describe the DL (object detection) models, baselines, and evaluation procedures used in our evaluations.

\begin{table}[t]
\caption{\textbf{Object Detection Models Used in Evaluation.} 
{\em Parameter counts are in millions; mAP@0.5:0.95.}}
\centering
\renewcommand{\arraystretch}{1.2}
\resizebox{1\columnwidth}{!}
{
\begin{tabular}
{l|c|c|c}
    \toprule
    \textbf{Model (Abbr.)} & \textbf{\# Params} & \textbf{Input Size} & \textbf{mAP} \\
    \midrule
    {\bf {YOLOv5-N}} (\yolo) & 1.9 M & 640 $\times$ 640  & 27.6  \\
    % (\textsc{Yolo}) & & & \\
    % \midrule
    {\bf FRCNN-MobileNetV3} (\frcnn) & 19.4 M & 640 $\times$ 640  & 32.8  \\
    % (\textsc{Frcnn}) & & & \\
    % \midrule
    {\bf RetinaNet-ResNet50} (\retina) & 38 M & 640 $\times$ 640  & 41.5  \\
    % (\textsc{RetinaNet}) & & & \\
    \bottomrule
\end{tabular}
}
\label{tab:tab-models}
\end{table}

\para{DL models}
We evaluate \sysname using three object detection models with varying computational demands (Table \ref{tab:tab-models}): YOLOv5-N (\yolo) with 1.9M parameters \cite{yolov5.web26}, FRCNN-MobileNetV3 (\frcnn) with 19.4M parameters \cite{mobilenetv3.iccv19, fasterrcnn-jieee}, and RetinaNet-ResNet50 (\retina) with 38M parameters \cite{retina-resnet-iccv, resnet.cvpr16}. 
This selection spans a 20$\times$ range in model size, allowing us to evaluate \sysname's effectiveness across lightweight to heavyweight inference workloads. 
The PyTorch framework is used to train these models on 80 object classes from the COCO dataset \cite{coco-dataset}. 
All models use an input size of 640 $\times$ 640. 
Among the three, \retina achieves the highest accuracy (41.5 of mAP), followed by \frcnn (32.8 of mAP) and \yolo (27.6 of mAP), consistent with their respective model size and capacities.

\para{Measurement Setup} 
We evaluate \sysname on two NVIDIA Jetson devices: \xavier and \orin (specifications in Table \ref{tab:devices} in \S\ref{subsec:background}). 
We limit CPU usage to 90\% using {\tt cgroups} to prevent resource contention. 
To collect system-level metrics, we use {\tt tegrastats}\footnote{\url{https://docs.nvidia.com/drive/drive-os-5.2.0.0L/drive-os/index.html}} (for power and GPU utilization/metrics) and \texttt{sysstat}\footnote{\url{https://github.com/sysstat/sysstat}} package (CPU and memory utilization/metrics). Measurements begin 2 seconds after inference starts and update every second, synchronized with throughput measurements.

\para{Configuration Space} 
We discretize the parameter ranges in Table \ref{tab:dvfs} (in \S\ref{subsec:background}) using practical step sizes: 100 MHz for CPU and GPU frequencies, and the available memory frequency options. 
This yields 2,160 configurations on \xavier (5 CPU cores $\times$ 8 CPU frequencies $\times$ 6 GPU frequencies $\times$ 3 memory frequencies $\times$ 3 concurrency levels) and 1,600 on Orin Nano (5$\times$8 $\times$4$\times$2$\times$5). 
Table \ref{tab:total-configs} shows the complete configuration space evaluated in our experiments. 
We exhaustively tested all parameter combinations and excluded only those that failed due to memory constraints or runtime errors. Heavier models have fewer valid configurations as they are more susceptible to these failures.

\begin{table}[t]
    \caption{\textbf{Evaluated Configuration Space per Model and Device.} {\em These numbers indicate the total number of configurations used in our evaluations.}}
    \centering
    \begin{tabular}
        {l|c|c}
        \toprule 
        {\bf Model} & \textbf{\xavier} & \textbf{\orin} \\
        \midrule
        \yolo & 2,067 & 1,522 \\
        \frcnn & 1,813 & 1,371 \\
        \retina & 1,491 & 1,223 \\
        \bottomrule
    \end{tabular}
    \label{tab:total-configs}
\end{table}

\para{Baseline Comparison} 
We compare \sysname against the following baselines:

\begin{itemize}
    \item {ORACLE}: 
    Exhaustive offline profiling of all configurations (the complete search space in Table \ref{tab:total-configs}) to identify the optimal setting within the evaluated search space. This serves as the upper-bound baseline.

    \item {ALERT} \cite{alert.atc20}:
    A profiling-based method using Kalman filtering to select configurations from offline profiling.

    \item {ALERT-ONLINE}: 
    An online version of ALERT replaces offline profiling with 10 random trials, matching \sysname's iteration budget for fair comparison.

    \item {\tt max-power}: The manufacturer's maximum performance preset, which sets all frequencies and core counts to their highest values.

    \item {\tt default}: The manufacturer's default power mode.

\end{itemize}

\para{Dataset} 
We use a one-second annotated real-world traffic-monitoring video as the input for inference. The DL model detects multiple objects in each frame. For video-based object detection, we first extract frames from the video using the \texttt{OpenCV}\footnote{\url{https://github.com/opencv/opencv}} library.

\begin{figure}[t]
  \centering
    \includegraphics[width=1\columnwidth]{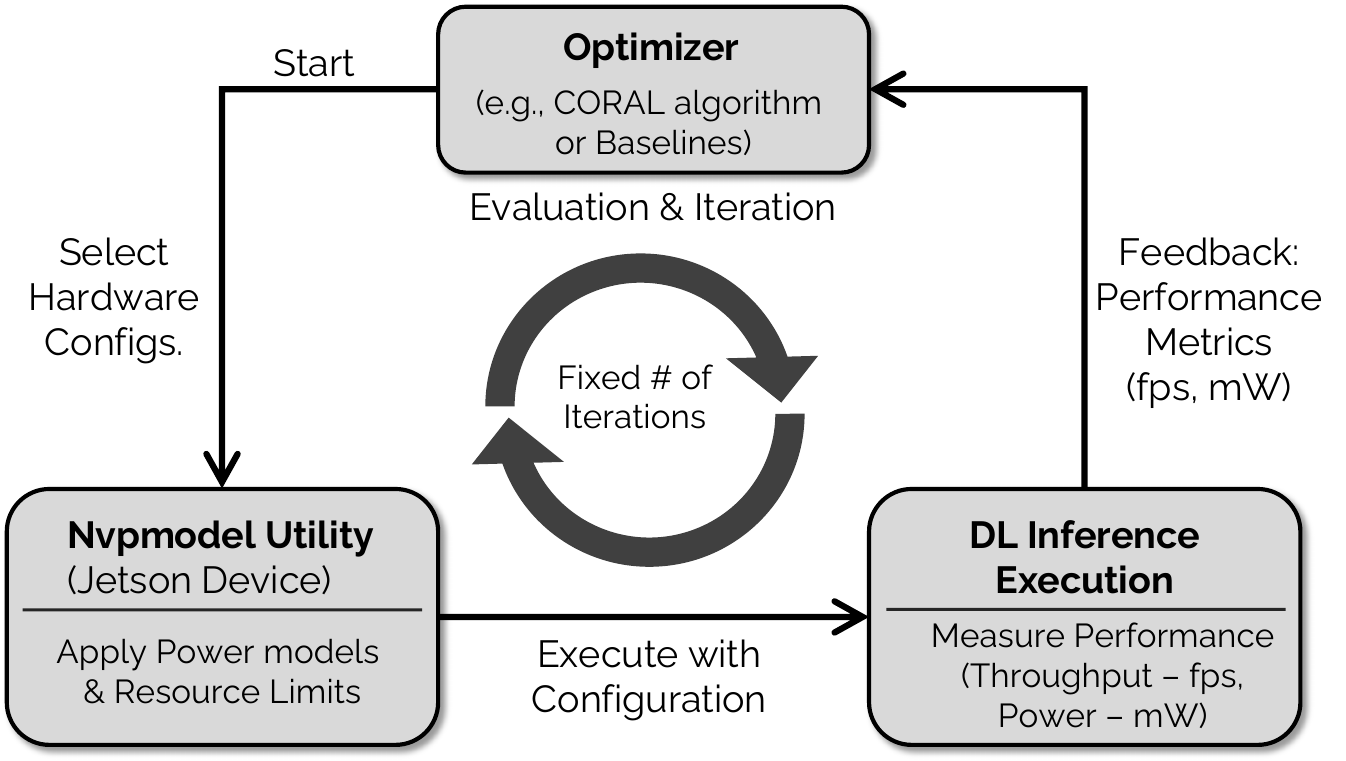}
  \caption{\textbf{Experimental Evaluation Loop for \sysname and Baselines.} \em }
    \label{fig:optimizer-diagram}
\end{figure}

\para{Evaluation Procedure} 
We implement all optimization methods within a common evaluation loop that applies configurations, executes inference, and collects performance metrics. The execution loop is illustrated in Figure~\ref{fig:optimizer-diagram}.

The evaluation process begins by initializing the optimizer, which can be either a baseline method or \sysname. 
The optimizer selects a hardware configuration, which is applied to the devices using the {\tt nvpmodel} utility \cite{nvpmodel}.
Once the configuration is set, the object detection inference is executed. 
The resulting performance metrics, throughput (fps) and power consumption (m\si{\watt}), are measured and fed back to the optimizer for evaluation. 
This feedback loop continues for a fixed number of iterations. We set the iteration budget to 10, which balances search effectiveness with practical deployment constraints. 
This budget is sufficient for \sysname to converge while remaining feasible for online adaptation scenarios.

\subsection{Evaluation Results}

We evaluate \sysname against the baselines in two scenarios: 1) a single-target scenario (optimizing for throughput only) and 2) a stricter dual-constraint scenario (optimizing for both simultaneously). 
All experiments use \yolo; results on larger models (\frcnn and \retina) are presented in \S\ref{subsec:generalization}.

\begin{figure}[t]
    \centering
    \begin{tabular}{cc}
    \includegraphics[width=0.47\linewidth]{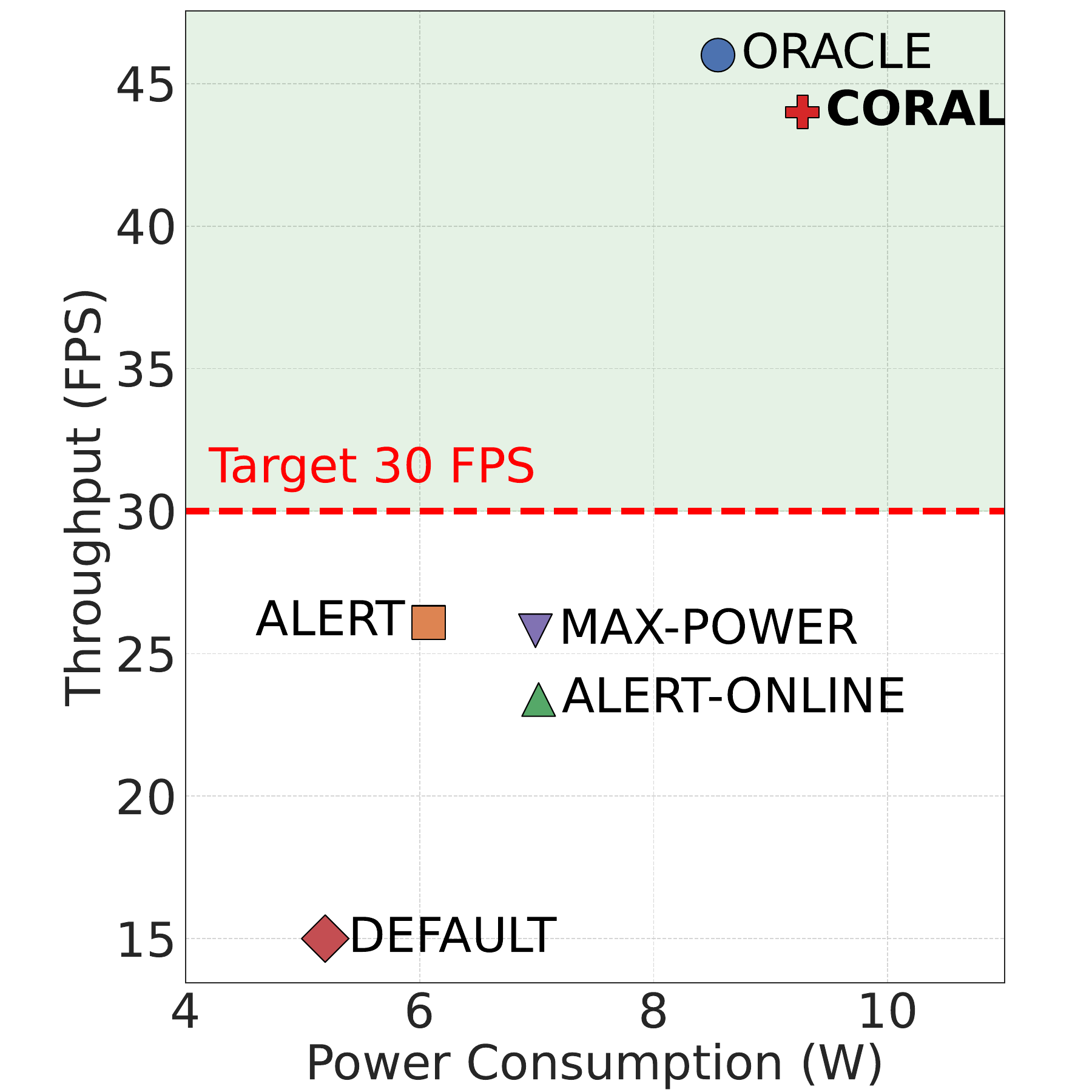} &
    \includegraphics[width=0.47\linewidth]{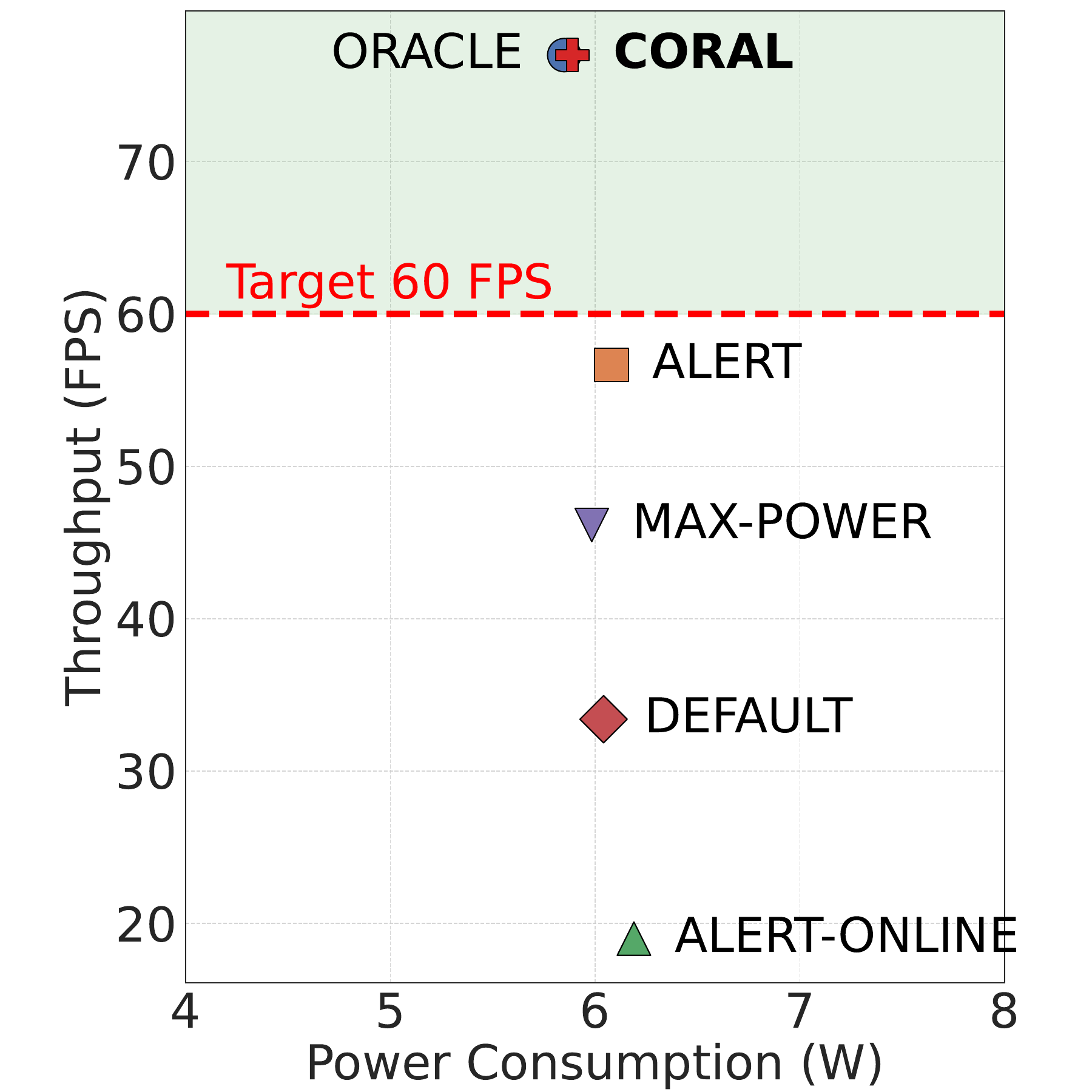} \\
    (a) \xavier  & (b) \orin 
    \end{tabular}
    \caption{{\bf Power-throughput trade-off in single-constraint scenario (\yolo).}}
    \label{fig:single_target_tradeoff}
\end{figure}

\begin{figure}[t]
    \centering
    \begin{tabular}{cc}
    \includegraphics[width=0.47\linewidth]{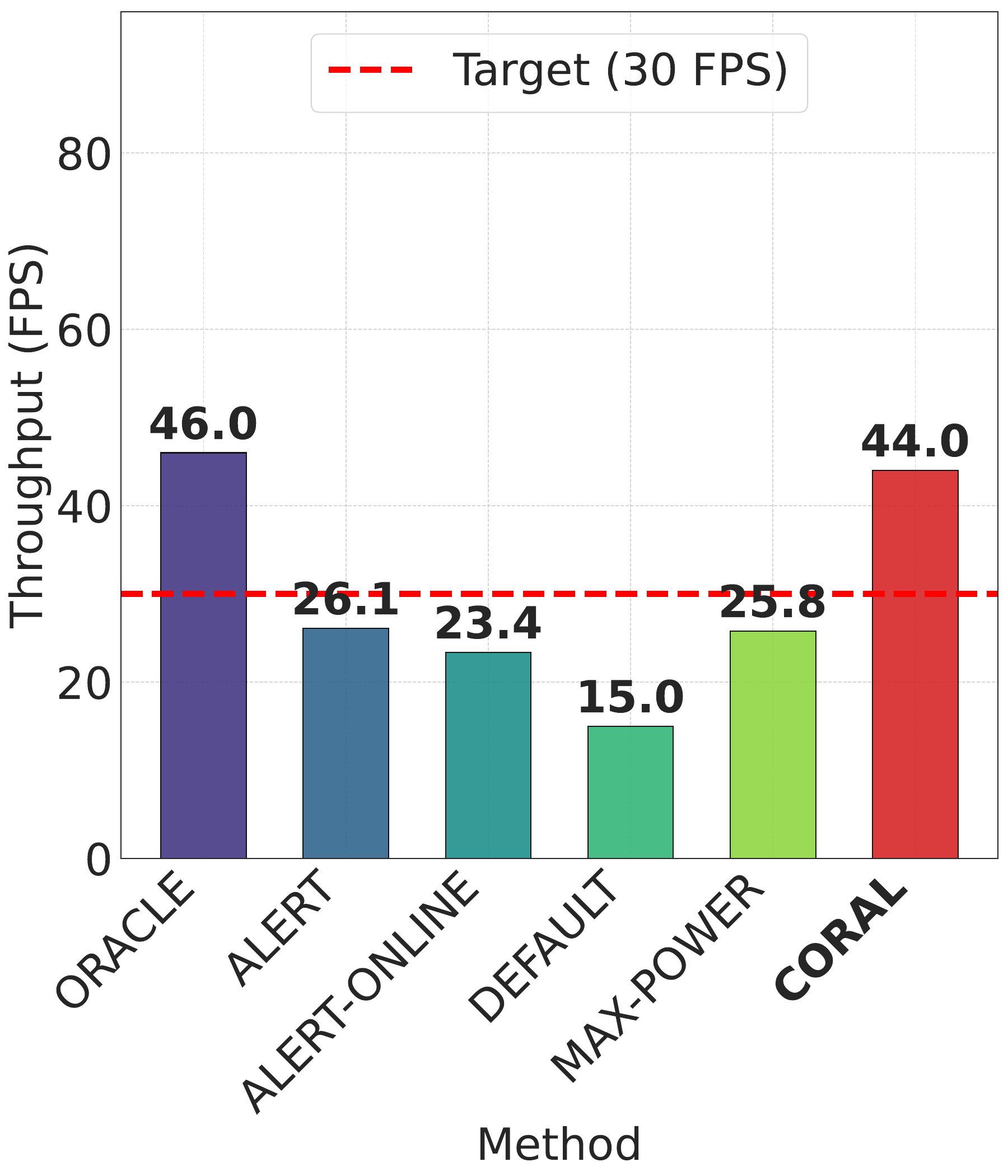} &
    \includegraphics[width=0.47\linewidth]{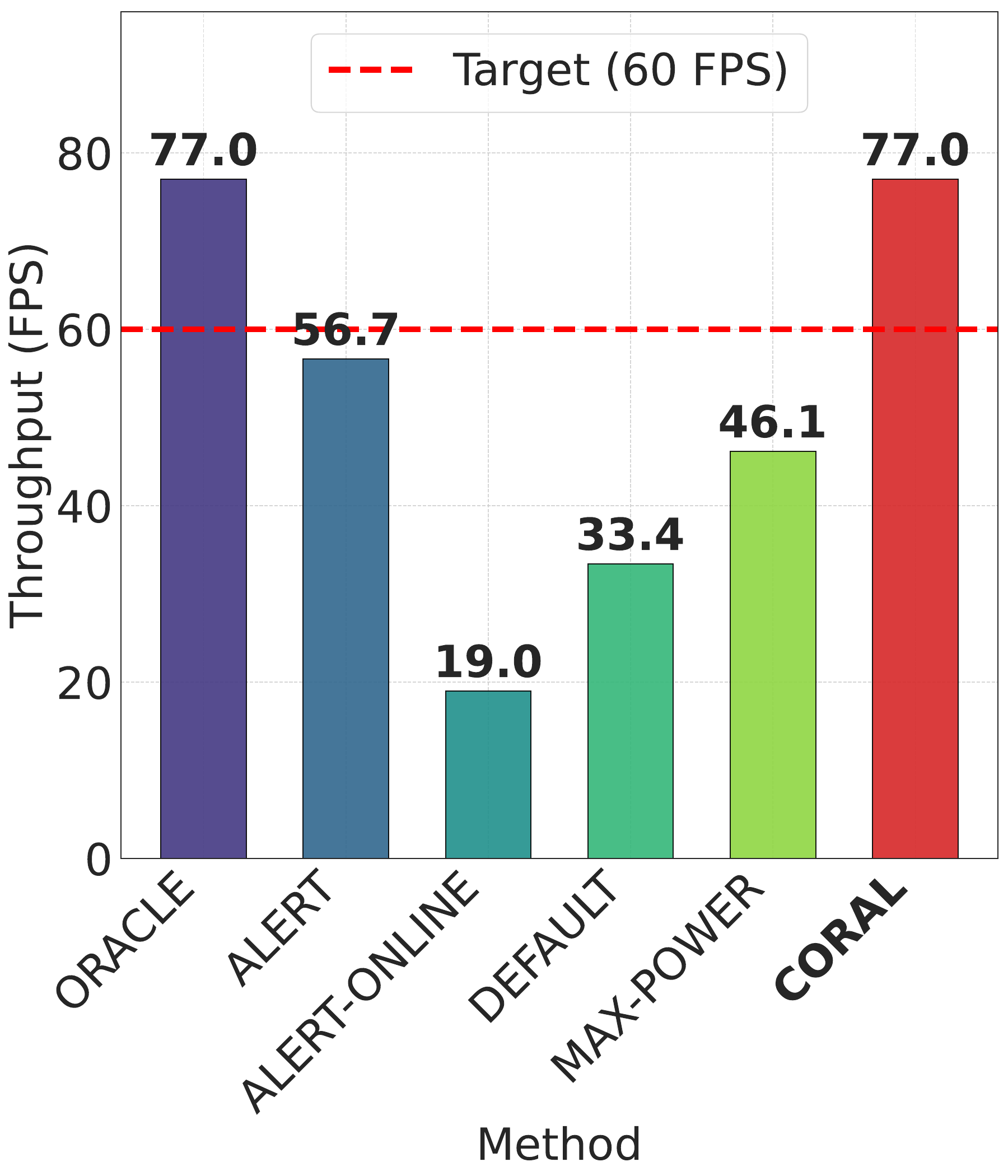} \\
    (a) \xavier & (b) \orin 
    \end{tabular}
    \caption{\bf Single-constraint (throughput) comparison results (\yolo).}
    \label{fig:single_target_throughput}
\end{figure}

\para{Single-Constraint Scenario} In the single-constraint scenario (Figures \ref{fig:single_target_tradeoff} and \ref{fig:single_target_throughput}), \sysname significantly outperforms manufacturer presets, achieving 96 -– 100\% of ORACLE throughput while presets ({\tt max-power} and {\tt default} modes) reach only 33\% –- 60\%. 
Figure \ref{fig:single_target_tradeoff} shows that presets either consume excessive power ({\tt max-power}) or sacrifice throughput ({\tt default}), whereas \sysname finds more balanced operating points. 
ALERT performs well in this scenario, leveraging its offline profiling data to filter optimal configurations, and slightly outperforms \sysname on \xavier. However, \sysname achieves these results without any offline profiling.

\begin{figure}[t]
    \centering
    \begin{tabular}{cc}
    \includegraphics[width=0.47\linewidth]{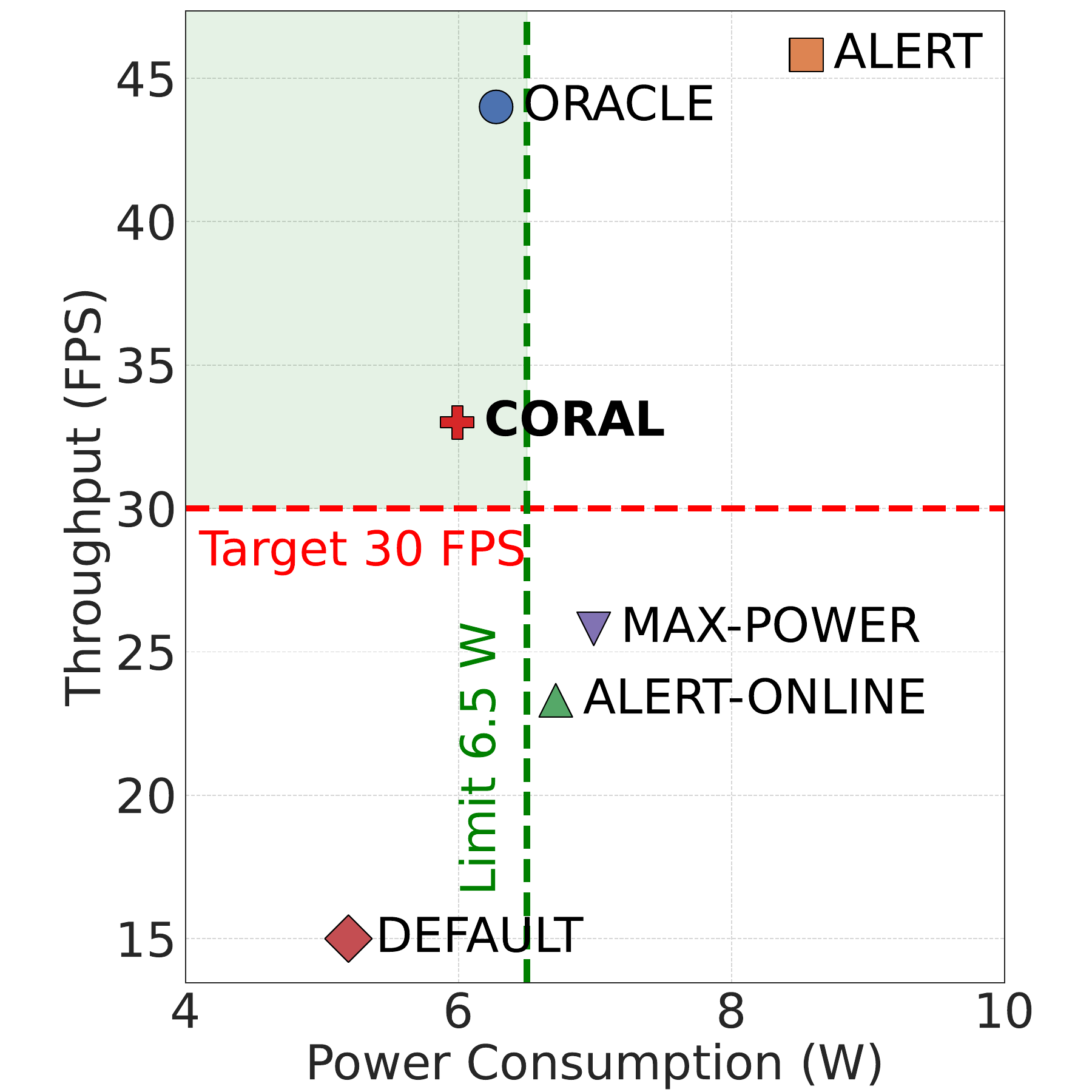} &
    \includegraphics[width=0.47\linewidth]{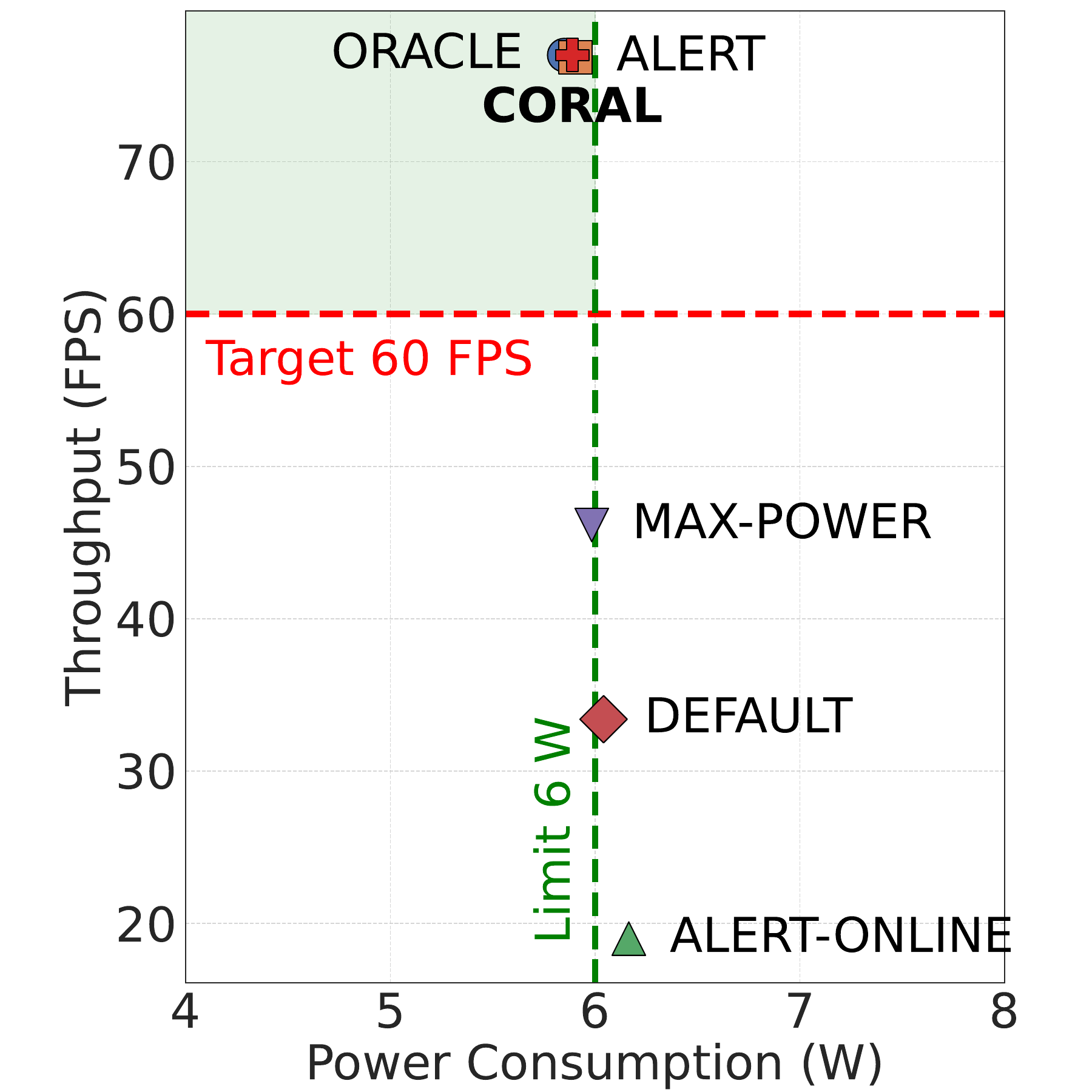} \\
    (a) \xavier  & (b) \orin 
    \end{tabular}
    \caption{{\bf Dual-constraint optimization results (\yolo).} 
    % {\em \sysname meets both power and throughput constraints where baselines fail.}
    }
    \label{fig:dual_target_tradeoff}
\end{figure}

\begin{figure}[t]
    \centering
    \begin{tabular}{cc}
    \includegraphics[width=0.47\linewidth]{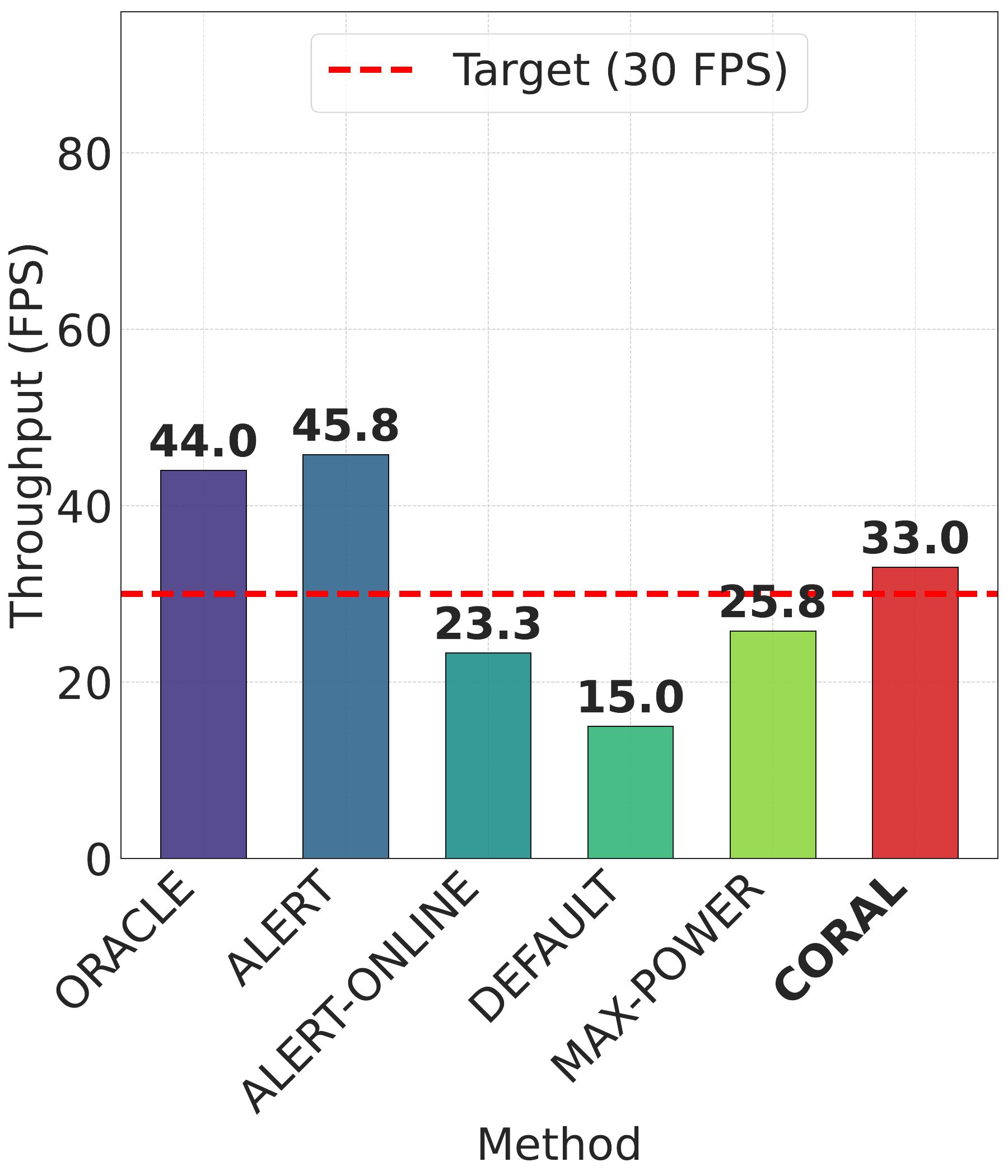} &
    \includegraphics[width=0.47\linewidth]{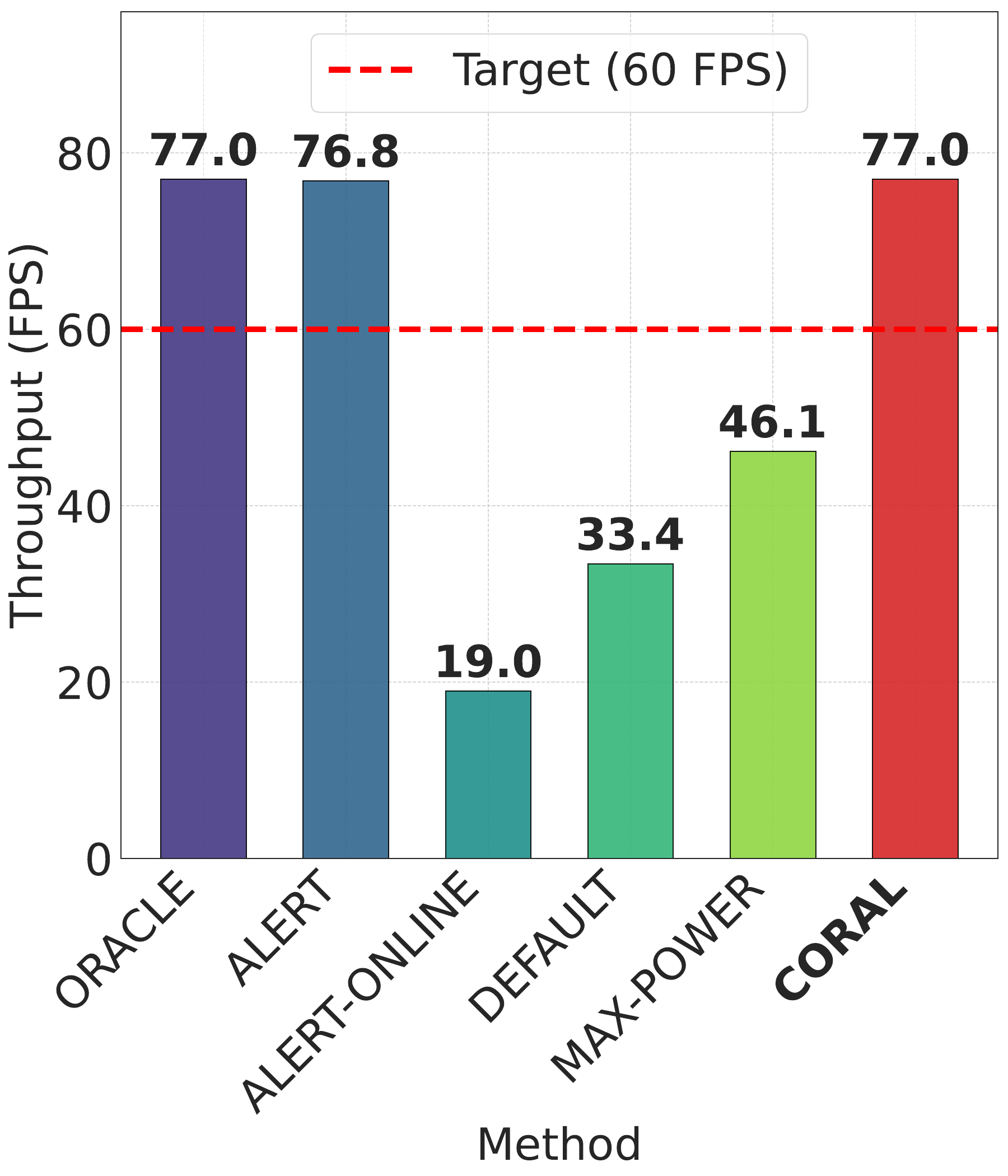} \\
    (a) \xavier  & (b) \orin 
    \end{tabular}
    \caption{{\bf Throughput achieved under dual-constraint scenarios (\yolo).}}
    \label{fig:dual_target_throughput}
\end{figure}

\para{Dual-Constraint Scenario}
The advantages of \sysname become evident in the dual-constraint scenario (Figures \ref{fig:dual_target_tradeoff} and \ref{fig:dual_target_throughput}). 
We set the following strict constraints:

\begin{itemize}
    \item \xavier: Power limit 6500 m\si{\watt}, Throughput target 30 fps.

    \item \orin: Power limit 5600 m\si{\watt}, Throughput target 60 fps.

\end{itemize}

In this evaluation, \sysname also successfully meets both constraints on both devices. 
On \xavier, \sysname achieves 33 fps at 5.5\si{\watt}, closely matching ORACLE (34 fps @ 5.9\si{\watt}). 
On \orin, \sysname achieves 77 fps at 5.8\si{\watt}, exceeding the throughput target while staying within the power budget.

In contrast, all baselines fail to satisfy the dual constraints.
ALERT prioritizes throughput and exceeds the power budget (8.5\si{\watt} on \xavier). 
ALERT-Online's random exploration fails to discover the narrow feasible region within the iteration budget. 
Manufacturer presets face a fundamental trade-off: {\tt max-power} violates power limits, while {\tt default} fails to meet throughput constraints.
\sysname's distance covariance-based search navigates the trade-off effectively without exhaustive offline profiling.

\subsection{Generalization Across Model Sizes}
\label{subsec:generalization}

The previous results use \yolo (with 1.9M parameters). 
Here, we test whether \sysname generalizes to models up to 20$\times$ larger: \frcnn (19.4M parameters) and \retina (38M parameters).
Consistent with the previous experiments, we evaluate under dual-constraint scenarios requiring both power limits and throughput targets to be satisfied.

\begin{figure}[t]
    \centering
    \begin{tabular}{cc}
    \includegraphics[width=0.47\linewidth]{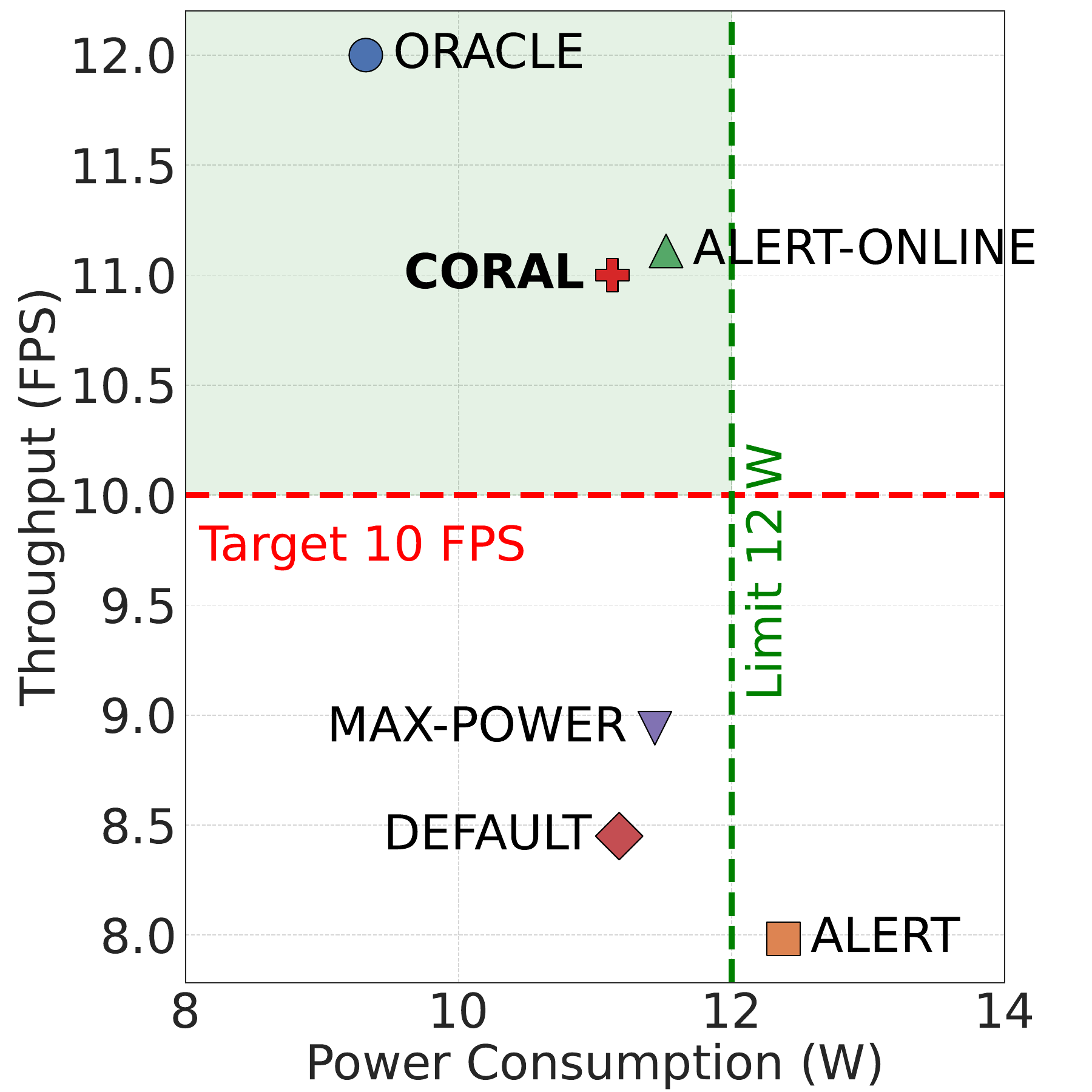} &
    \includegraphics[width=0.47\linewidth]{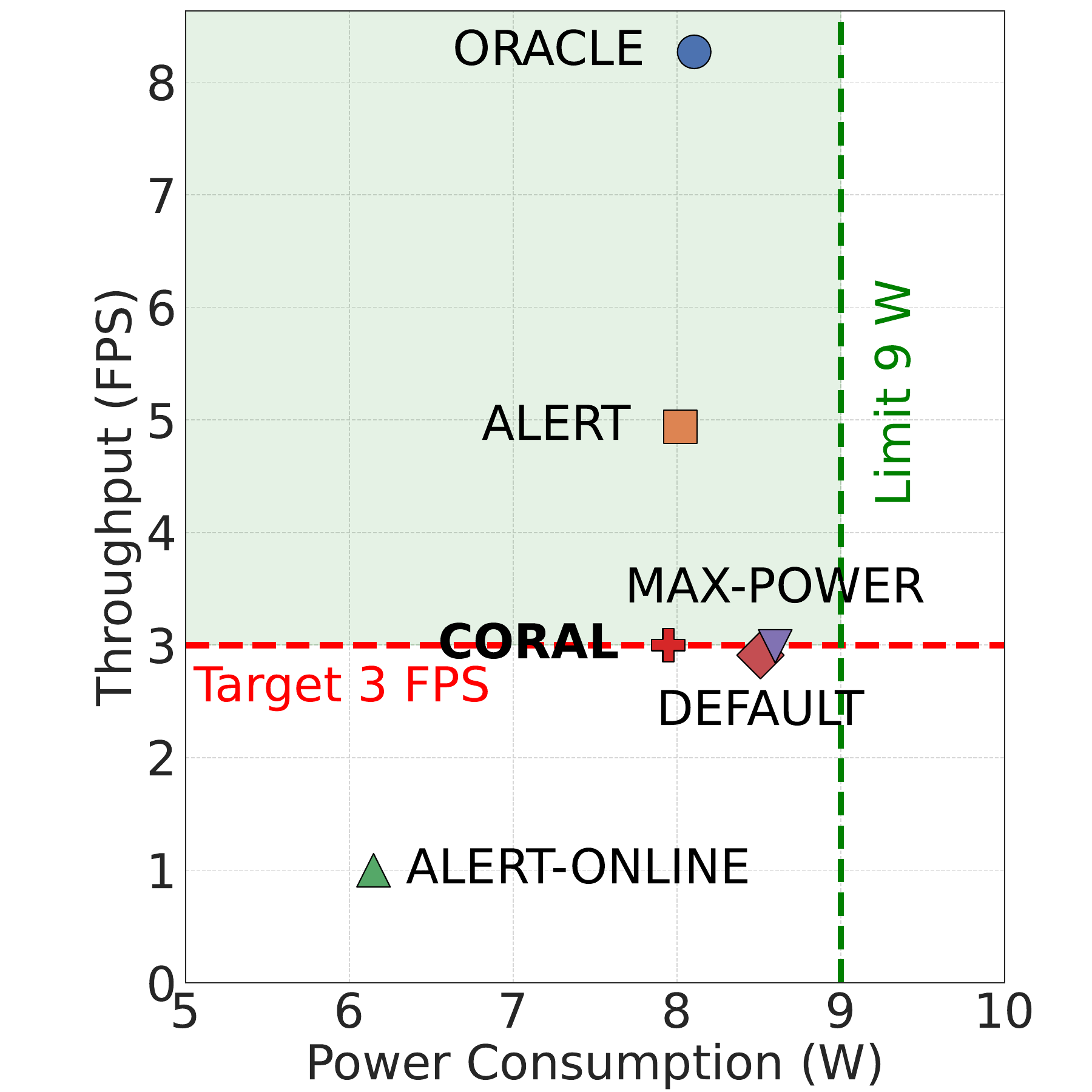} \\
    (a) \xavier  & (b) \orin 
    \end{tabular}
    \caption{{\bf Power-throughput trade-off on \frcnn model under dual-constraint scenarios.}}
    \label{fig:frcnn_tradeoff}
\end{figure}

\begin{figure}[t]
    \centering
    \begin{tabular}{cc}
    \includegraphics[width=0.47\linewidth]{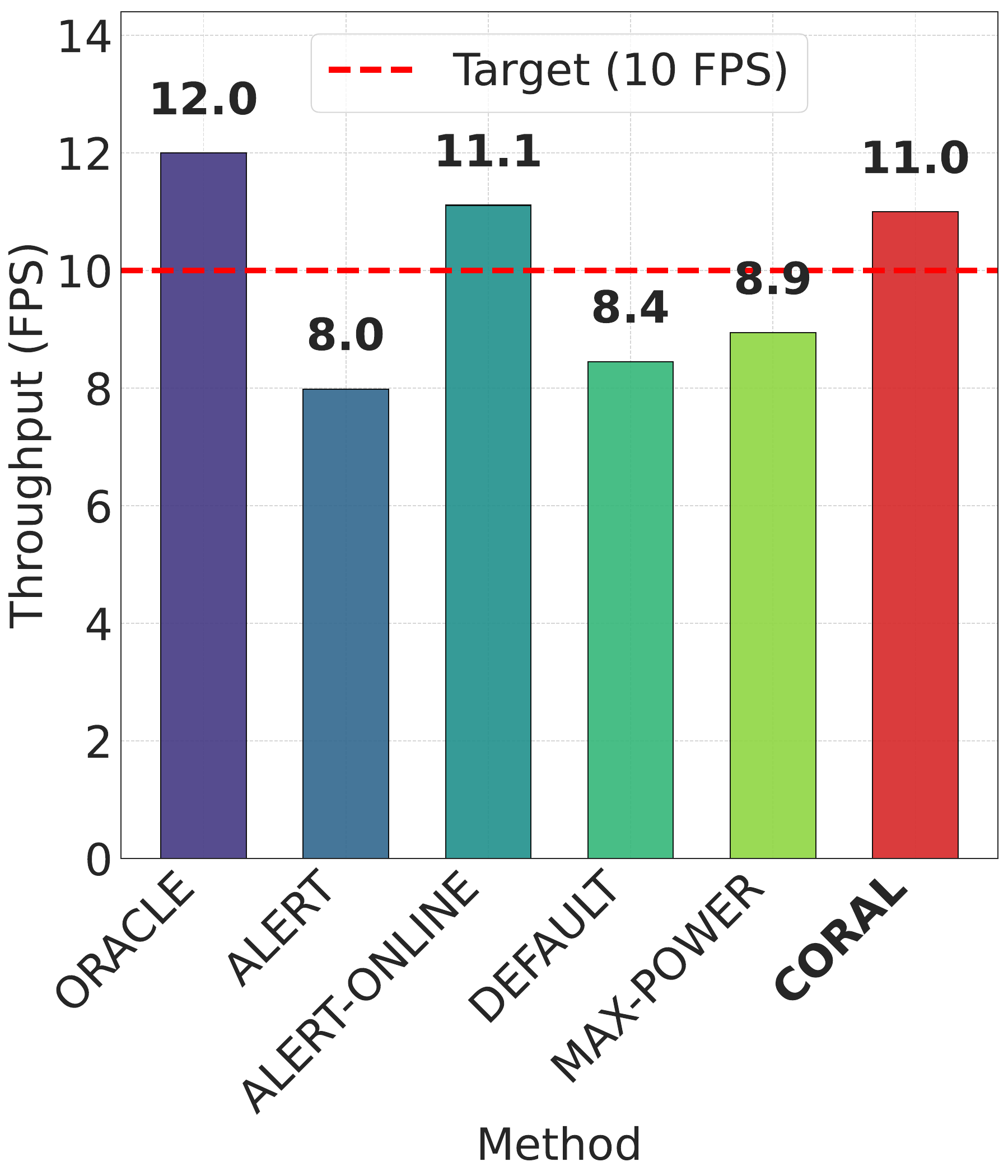} &
    \includegraphics[width=0.47\linewidth]{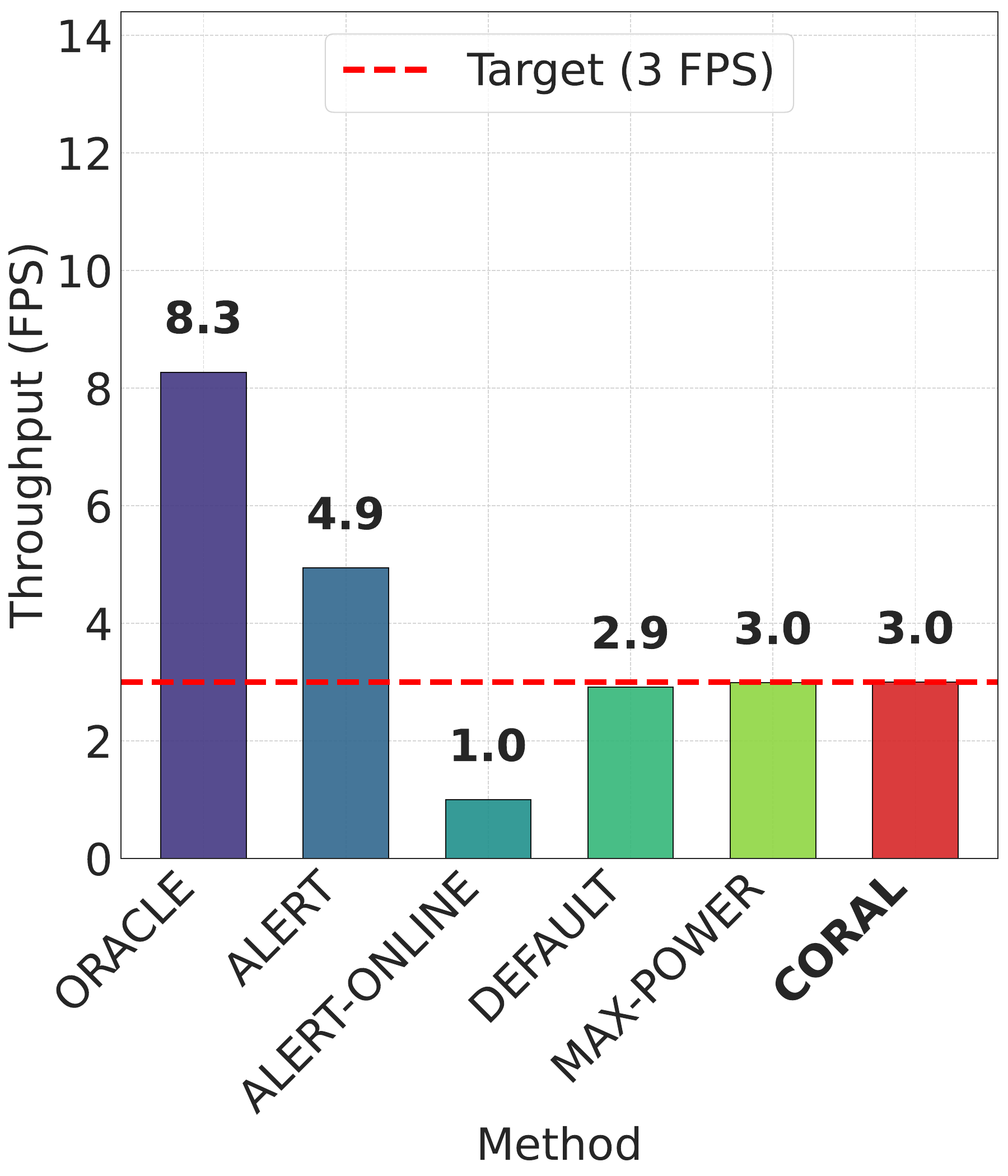} \\
    (a) \xavier  & (b) \orin 
    \end{tabular}
    \caption{{\bf Throughput comparison on \frcnn model under dual-constraint scenarios.}}
    \label{fig:frcnn_throughput}
\end{figure}

\begin{figure}[t]
    \centering
    \begin{tabular}{cc}
    \includegraphics[width=0.47\linewidth]{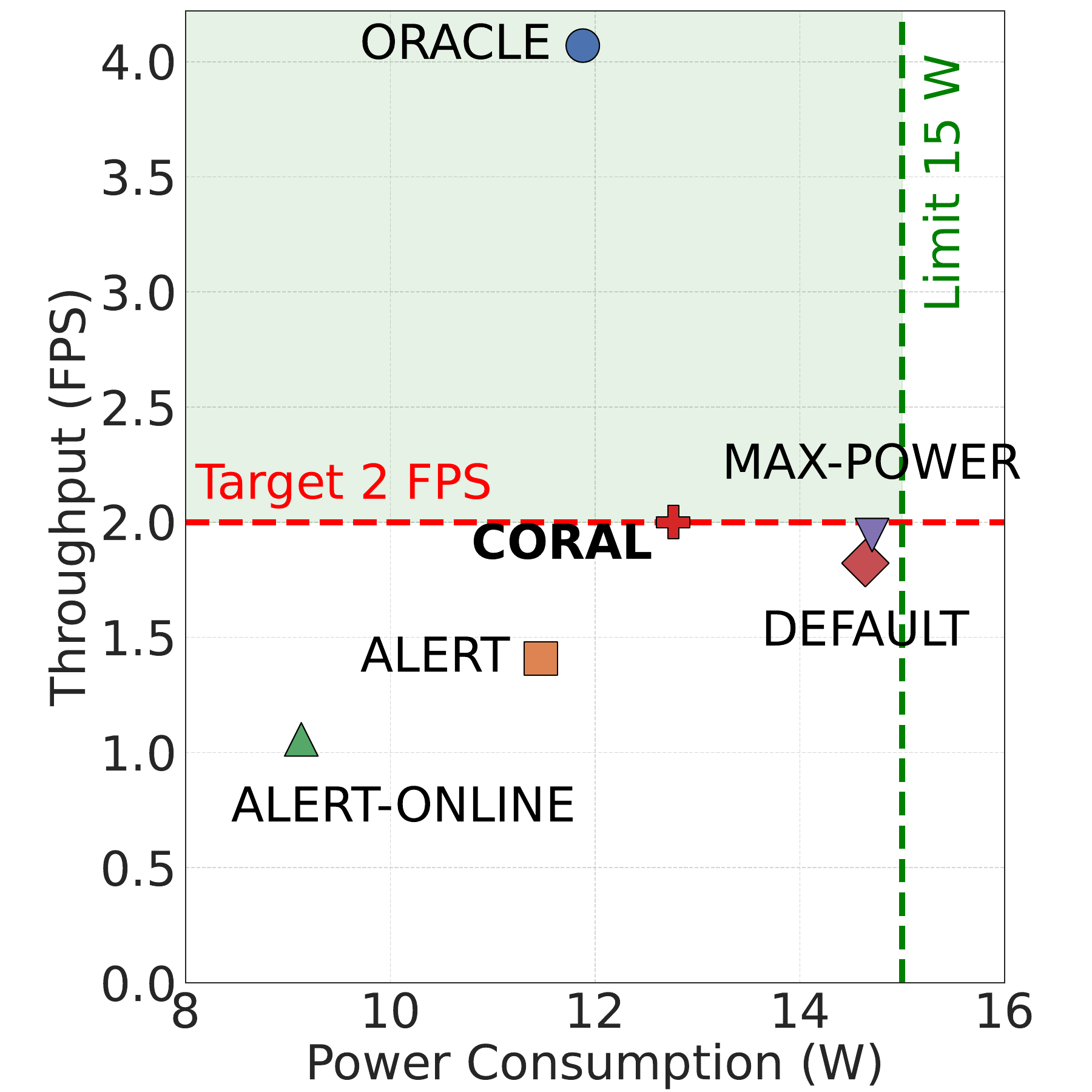} &
    \includegraphics[width=0.47\linewidth]{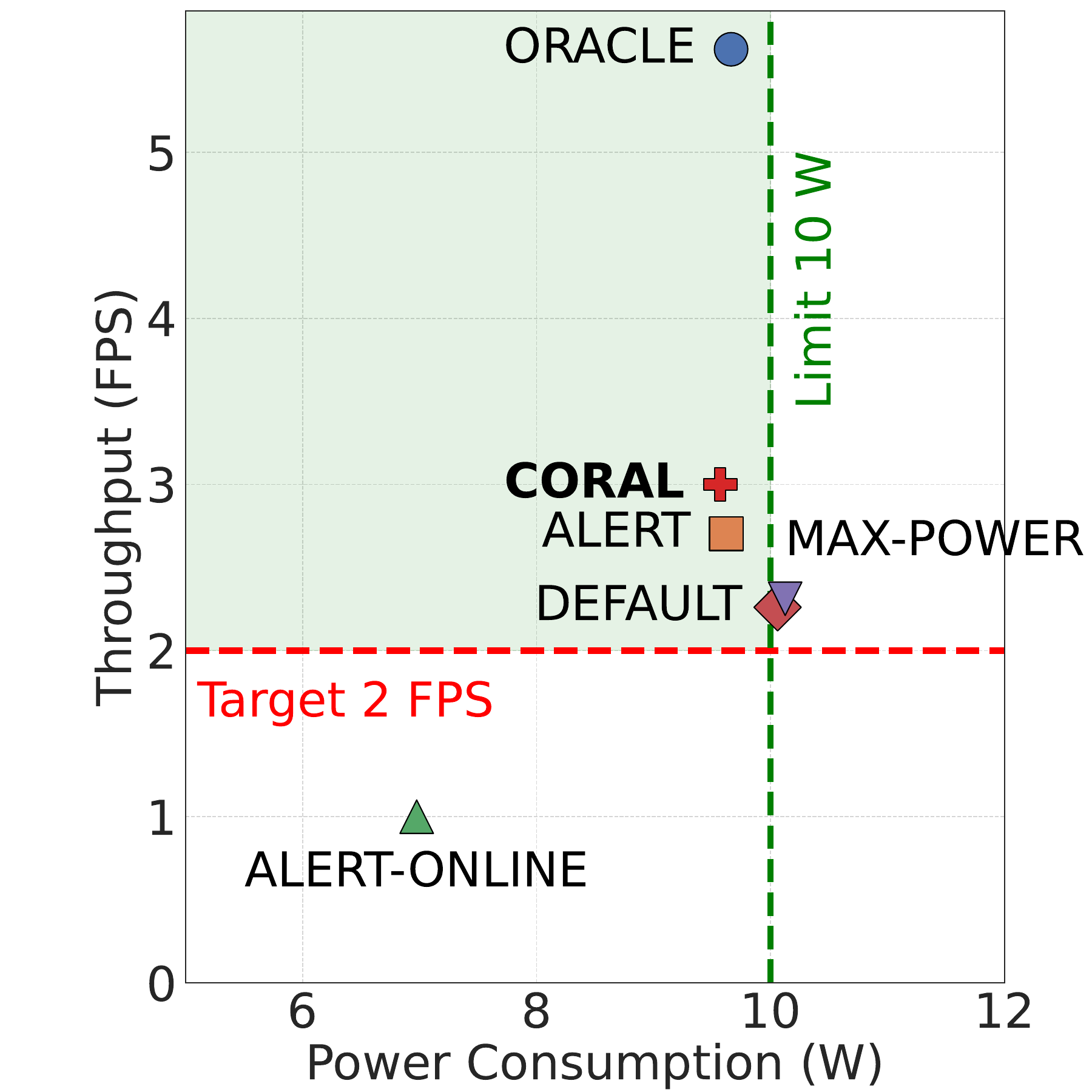} \\
    (a) \xavier  & (b) \orin 
    \end{tabular}
    \caption{{\bf Power-throughput trade-off on \retina model under dual-constraint scenarios.}}
    \label{fig:retinanet_efficiency}
\end{figure}

\begin{figure}[t]
    \centering
    \begin{tabular}{cc}
    \includegraphics[width=0.47\linewidth]{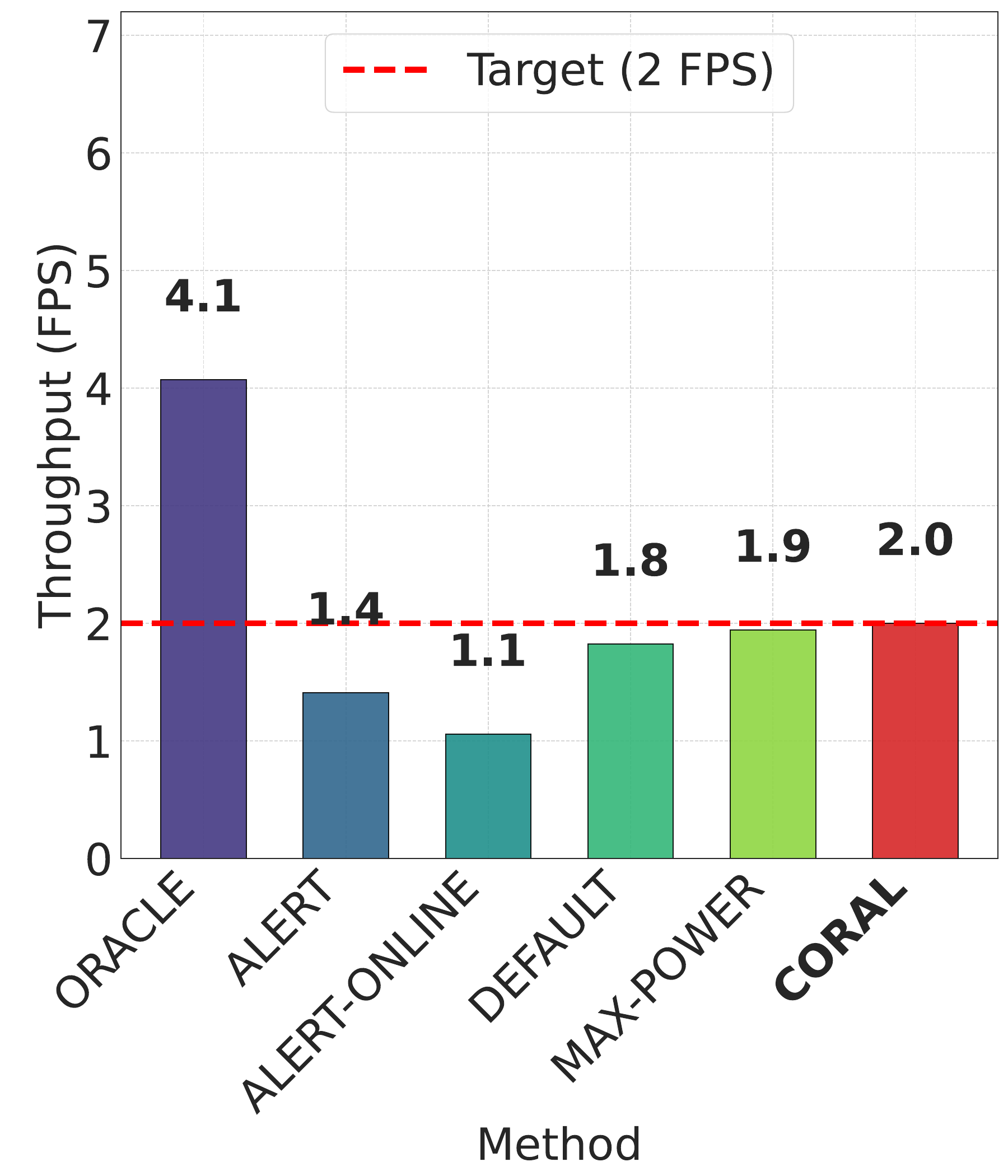} &
    \includegraphics[width=0.47\linewidth]{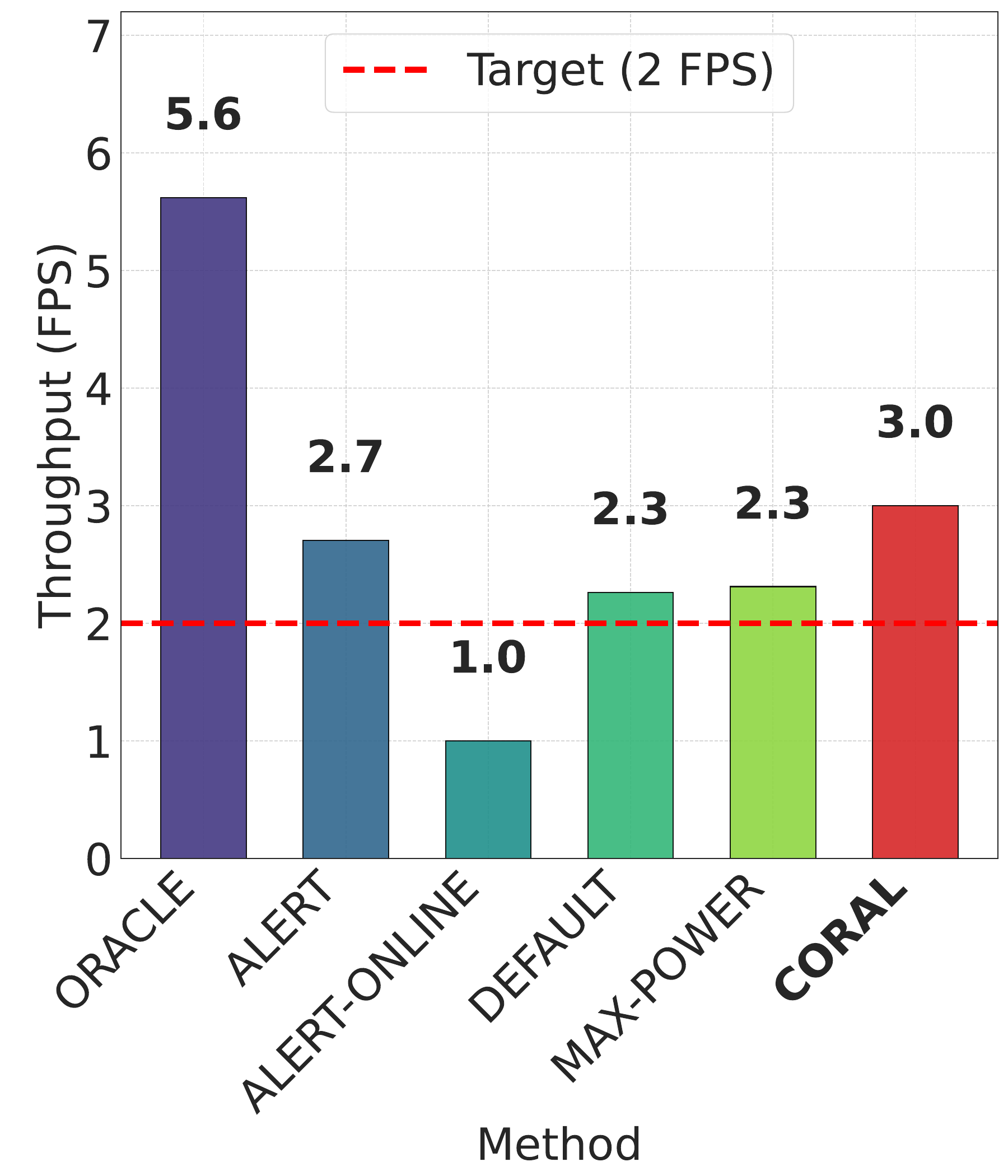} \\
    (a) \xavier  & (b) \orin 
    \end{tabular}
    \caption{{\bf Throughput comparison on \retina under dual-constraint scenarios.}}
    \label{fig:retinanet_throughput}
\end{figure}

The results reveal an interesting trend. 
As models grow larger, baseline methods struggle more, while \sysname maintains near-ORACLE performance.
For \frcnn (Figures \ref{fig:frcnn_tradeoff} and \ref{fig:frcnn_throughput}), \sysname achieves both constraints (10 fps / 12\si{\watt} on \xavier) while ALERT exceeds the power budget. 
For \retina (Figures \ref{fig:retinanet_efficiency} and \ref{fig:retinanet_throughput}), the gap grows further. 
\sysname matches ORACLE exactly, whereas all baselines fail to find valid configurations.

This growing gap suggests \sysname becomes increasingly valuable as constraints tighten. 
Heavier models leave less headroom for suboptimal configurations, making \sysname's targeted search essential for finding the narrow feasible region.

\section{Conclusion}\label{sec:conclusion}

We presented \sysname, an online optimization method for deep learning inference on edge devices. 
\sysname addresses the throughput-power co-optimization problem, \eg, satisfying both throughput targets and power budgets simultaneously, {\em without requiring offline profiling}. 
By leveraging distance covariance, it captures non-linear dependencies between hardware parameters and performance metrics, enabling near-optimal configurations quickly and with minimal overhead.

Our evaluation on two edge devices across three object detection models (\yolo, \frcnn, and \retina) highlights \sysname's effectiveness. 
In single-constraint scenarios, \sysname achieves 96\% -- 100\% of the optimal throughput found by offline exhaustive search. 
In dual-constraint scenarios with strict power and throughput constraints, \sysname consistently finds right configurations while baseline methods either exceed power budgets or fail to meet throughput targets.

\bibliographystyle{unsrt}
\bibliography{bibfiles/reference.bib}

\end{document}